\documentclass[a4paper,usenatbib,useAMS]{mnras}

\usepackage{graphicx}
\usepackage{lscape}
\usepackage{perpage}
\usepackage{amsmath}
\usepackage{color}
\MakePerPage{footnote}

\newcommand{\beq}	{\begin{equation}}
\newcommand{\eeq}	{\end{equation}}
\newcommand{\beqa}	{\begin{eqnarray}}
\newcommand{\eeqa}	{\end{eqnarray}}
\newcommand{\e}	        {$^{-1}$}
\newcommand{\ee}	{$^{-2}$}
\newcommand{\eee}	{$^{-3}$}
\newcommand{\dis}		{\displaystyle}

\newcommand{\calm}	{{\cal M}}

\newcommand{\caln}      {{\cal N}}

\newcommand{\avg}[1]    {{\langle #1 \rangle}}

\newcommand\fnm		{\footnotemark}
\newcommand\fnt		{\footnotetext}

\newcommand{\alfven}    {{Alfv$\acute{\rm e}$n }}

\newcommand{\muphi}	{\mu_{\Phi}}

\newcommand{\va}	{v_{\rm A}}

\newcommand{\vrms}	{v_{\rm rms}}

\newcommand{\avir}      {\alpha_{\rm vir}}

\newcommand\brms        {B_{\rm rms}}

\newcommand{\eff}       {\epsilon_{\rm ff}}

\newcommand{\ma}	{{\calm_{\rm A}}}
\newcommand{\mao}	{{\calm_{\rm A,0}}}

\newcommand{\mmug}	{\mu{\rm G}}
\newcommand{\nbh}	{\bar n_{\rm H}}

\newcommand{\nbht}	{\bar n_{\rm H,\, 3}}

\newcommand{\sid}	{\sigma_{\rm 1D}}
\newcommand{\snt}       {\sigma_{\rm nt}}
\newcommand{\spc}	{\sigma_{\rm pc}}
\newcommand{\spcs}      {{\sigma_{\rm pc}^*}}

\newcommand{\NH}	{N_{\rm H}}
\newcommand{\tff}	{t_{\rm ff}}
\newcommand{\tffo}	{t_{\rm ff,0}}

\newcommand\cs		{c_{\rm s}}
\def\eff				{\epsilon_{\rm ff}}
\def\efft				{\epsilon_{\rm ff,\,T}}

\newcommand\pc		{{\rm pc}}
\def\rhomm			{\avg{\rho}_M}
\def\tff				{t_{\rm ff}}
\def\tsf				{t_{\rm SF}}

\title{The Formation of Stellar Clusters in Magnetized, Filamentary Infrared Dark Clouds}
\author[Pak Shing Li, Richard I. Klein, and Christopher F. McKee]{Pak Shing Li$^{1}$\thanks{E-mail:psli@astron.berkeley.edu (PSL)}, Richard I. Klein$^{1,3}$\thanks{E-mail:klein@astron.berkeley.edu (RIK)}, and Christopher F. McKee$^{1,2}$\thanks{E-mail:cmckee@astro.berkeley.edu (CFM)}\\
$^{1}$Astronomy Department, University of California, Berkeley, CA 94720\\
$^{2}$Physics Department, University of California, Berkeley, CA 9472of0\\
$^{3}$Lawrence Livermore National Laboratory,P.O.Box 808, L-23, Livermore, CA 94550}
\begin{document}

\date{}

\pubyear{}

\label{firstpage}
\pagerange{\pageref{firstpage}--\pageref{lastpage}}
\maketitle

\begin{abstract}
Star formation in a filamentary infrared dark cloud (IRDC) is simulated over a dynamic range of 4.2 pc to 28 au for a period of $3.5\times 10^5$ yr,
including magnetic fields and both radiative and outflow feedback from the protostars.
At the end of the simulation, the star formation efficiency is 4.3 per cent and the star formation rate per free fall time is $\eff\simeq 0.04$, within the range of observed values \citep{kru12a}. The total stellar mass increases as $\sim\,t^2$, whereas the number of protostars increases as $\sim\,t^{1.5}$. We find that the density profile around most of the simulated protostars is $\sim\,\rho\propto r^{-1.5}$, as predicted by \citet{mur15}.
At the end of the simulation, the protostellar mass function  
approaches the \citet{chab05} stellar initial mass function.
We infer that the time to form a star of median mass $0.2\,M_\odot$ is about $1.4\times 10^5$~yr from the median mass accretion rate.
We find good agreement among the protostellar luminosities observed in the large sample
of \citet{dun13}, our simulation, and a theoretical estimate, and conclude that the classical protostellar luminosity problem \citep{ken90} is resolved. The multiplicity of the stellar systems in the simulation agrees to within a factor 2 of observations of Class I young stellar objects; most of the simulated multiple systems are 
unbound.
Bipolar protostellar outflows are launched using a sub-grid model, and extend up to 1 pc from their host star. The mass-velocity relation of the simulated outflows is consistent with both observation and theory.
\end{abstract}


\section{Introduction}

Large dark cloud filaments are commonly observed in Giant Molecular Clouds (GMCs) 
over a wide range of length scales
\citep[see reviews by][and references therein]{ber07,and14}. Many of these filamentary dark clouds have column densities higher than $10^{22}$ cm$^{-2}$ and volume densities higher than $10^4$ cm$^{-3}$. Such clouds are called Infrared Dark Clouds (IRDCs) because of their high
 extinction at mid-IR \citep{par96,car98,ega98} or even far-IR wavelengths against the galactic background; an example of the latter is G028.37+00.07 \citep[e.g.][]{lim14}.
The masses of IRDCs are generally of order hundreds to thousands of solar masses, and they are believed to be the precursors of massive stars and star clusters formation. Observations of long filamentary clouds show that they
often have a mass per unit length significantly greater than the maximum that can be supported against gravitational collapse by thermal pressure,  $2c_s^2/G$ (e.g., \citealp{inu97}),
which is $16.6\, M_{\odot}/{\rm pc}$ at 10 K; as a result,
in the absence of sufficient turbulent or magnetic support, they are undergoing global gravitational contraction \citep[e.g.][]{and10,pol13,kon15}.
Sub-millimeter observations of IRDCs \citep[e.g][]{kon10,wan14,zha15} reveal hierarchical fragmentation that leads to dense cloud clumps and cores on $0.01 \sim 1$ pc size scales inside the filaments. 
IRDCs  typically have temperatures of order $10 \sim 15$ K, but the temperature of some of the dense cores 
is higher, indicating the possible formation of protostars at an early stage (e.g., \citealp{zha15}). Water masers and outflows are also detected in some of the dense clumps in IRDCs. For example, \citet{wan06} found that about 14 per cent of the IRDC clumps in their sample have water masers, and \citet{zha15} detected outflows in CO 2-1 and SiO 5-4 from the dense cores
in IRDC G28.34+0.06 P1.

It is natural to expect that star clusters will eventually form in  
IRDCs, particularly the more massive ones
However, the evolutionary process of cluster formation is unclear because what we have are individual observational snapshots from different 
IRDCs. Most IRDCs are filamentary, and we shall focus on them.
Several questions present themselves: Would stars form throughout the filament, or would the formation propagate along the filament? How efficient is protostellar feedback in destroying or disrupting a filament? Would the geometry of long filamentary clouds affect the properties of a protostellar cluster, such as the protostellar mass function (PMF) and the companion multiplicity?
To address these questions, high-resolution, large-scale turbulence simulations with multi-physics are crucial in providing a detailed picture of cluster formation inside filamentary clouds.
Magnetic fields are found to be important in dense clumps and cores; for example,
\citet{zha14} measured the polarization angles of clumps and cores and found a bimodal distribution within $40^{\circ}$ and around $90^{\circ}$
relative to the parsec-scale magnetic field,
indicating that the magnetic field orientation within clumps and cores is not random but organized.
Our recent detailed analysis of magnetized cloud clumps in IRDC simulations \citep{li15} demonstrates the important role of the magnetic field and supersonic turbulence in the clump evolution process. In fact, our simulations show that a moderately strong magnetic field is also important in
forming long filamentary clouds within the larger clouds.
 Our study of the formation, structure, and dynamics of filamentary clouds will be reported in a separate paper (Li et al. 2017, in preparation). 

In this paper, we present the results from a zoom-in adaptive mesh refinement (AMR) simulation based on our recent large-scale, ideal MHD, isothermal simulation of a turbulent box, which had a size of 4.55 pc for fiducial values of the parameters \citet{li15}.
The advantage of using this large-scale simulation as a starting point is that the zoom-in simulation begins with a filamentary cloud that has already formed through the natural MHD turbulent cascade from the large-scale turbulent driving and through gravitational collapse, instead of beginning with some ad-hoc initial conditions. With the knowledge of where the filamentary cloud forms, we focus on a 4.2 pc$^3$ region surrounding a long massive filament and follow the gravitational collapse down to 28 au resolution in order to study the properties of the stellar cluster formed within the filament. In this zoom-in simulation, we have continuous turbulence driving at all times. Radiative transfer is included to understand the effect of radiation feedback in the formation of a protostellar cluster. In Section \ref{sec:sim}, we describe our methodology and the model parameters of the zoom-in simulation. In Section \ref{sec:results}, we present the results of the zoom-in simulation on the properties of the protostellar cluster and compare with observations, including the temperature evolution of the cluster, 
core accretion and star formation rates, the PMF and protostellar luminosity function (PLF) of the star cluster, the protostellar multiplicity, and the protostellar outflow properties. The conclusions are summarized in Section \ref{sec:conclusion}.

\section{Simulations}
\label{sec:sim}

\subsection{Numerical Methods}
\label{method}

Our protostellar cluster simulations are performed using the \textsc{orion2} adaptive mesh refinement (AMR) code which solves   for MHD \citep{gar05,mig07,sto08,li12} along with coupled self-gravity \citep{mar08}
and radiation transfer \citep{kru07} in the two-temperature, mixed-frame,
grey, flux-limited diffusion approximation.  \textsc{orion2} utilizes a conservative second order Godunov scheme to solve the equations of compressible gas dynamics \citep{tru98,kle99}. 
The exact set of equations
solved are the same as those given in \citet{mye13}.  Our
radiative transfer calculations use the frequency-integrated grey dust
opacities from the iron-normal, composite aggregates model of \citet{sem03}.
 
This present work is based on a large-scale AMR simulation \citep{li15}
to study the structure and formation of Infrared Dark Clouds (IRDCs). The large-scale simulation uses our adaptive mesh refinement (AMR) radiation-magnetohydrodynamic code \textsc{orion2} (cf. references above) with ideal MHD and self-gravity to follow the evolution of driven turbulence on a base grid of $512^3$ with two levels of grid refinement. Periodic boundary conditions are adopted and the system is driven for two crossing times with no self-gravity in order to produce self-consistent turbulent density and velocity fields. The first crossing time is at the base level and the second has two levels of refinement. The refinement thresholds for pressure jumps (thermal pressure + magnetic pressure) and shear flows are chosen so that at least $12-15$ per cent of the volume is refined at all times during the second crossing time. This ensures that before gravity is turned on, the high-mass end of the probability density function (PDF) of volume density is equivalent to what would be obtained in a $2048^3$ single-grid simulation \citep{li12}.  The driving is at the largest scale, with wave numbers $k = 1-2$, and it uses the \citet{mac99} recipe.   After two crossing times, we turn on gravity and set $t = 0$ at this moment.  The driving continues with a constant energy injection rate that maintains the system globally at a constant sonic Mach number, $\calm = 10$.  Therefore, the simulation maintains continuously driven turbulence.

Dense regions inside the system undergo gravitational collapse, so it is necessary to add the Jeans condition for refinement \citep{tru97}.
We initialize a sink particle in any zone
on the finest AMR level that becomes dense enough 
to reach a local Jeans number of 1/8, so that the Jeans length is resolved by 8 cells.  Sink particles
then evolve and interact with 
other sink particles and the gas
through gravity according
to the methodology of \citet{kru04}, updated to include the
effects of magnetic fields on the rate of gas accretion onto the sink
particles \cite[see the appendix of][]{lee14}. 
Two sink particles within the accretion zone ($4\Delta x\simeq 112$ au) are merged if the smaller one has a mass less than $0.01 M_{\odot}$ in order to prevent the creation of a large number of very small particles.

The sink particles include source terms 
in the momentum and energy equations
to capture the effects of
protostellar feedback that originate on smaller length scales than
those resolved in the model. The sink particles thus become star particles that emit radiation
and heat the ambient medium \citep{kru07} and that have protostellar outflows \citep{cun11}. 
We determine the luminosity using a simple one-zone protostellar evolution model introduced by \citet{nak95} and extended by \citet{nak00} and \citet{tan04}.  We use the updated version of this model described by \citet{off09}
in which a series of different phases of the star’s evolution are treated starting from pre-collapse to deuterium burning and onto the main sequence (see the appendix in \citet{off09}).
The protostellar radius is calculated based a version of the \citet{mck03}/\citet{tan04} model modified to agree with the results
in Appendix C of \citet{hos09}. Since the \citet{hos09} results do not extend below $0.1\,M_\odot$, the protostellar radius is more 
uncertain there.
Following \cite{off09}, the luminosity that is imparted to the star particle consists of components from the stellar interior and the accretion shock, with 3/4 of the energy radiated away and the rest carried away by a wind 
from the disk associated with the protostar.   
A correction to the mean accretion luminosity of a protostar for the ionization and dissociation energies was introduced in the calculation of radiative feedback in the simulation after the simulation was completed (see Section 3.5.1).

We include feedback from protostellar outflows around each sink
particle by following the procedure described in \citet{cun11}, who set 
the outflow mass ejection rate by two parameters: the fraction of accreted gas
that is ejected into the outflow, $f_w$, and the outflow ejection speed,
$v_w$.  However, in the present work we use as our model parameter choices those of 
\citet{han12} by setting $f_w=0.3$ and $v_w = v_{\rm Kep}/3$,
where $v_{\rm Kep}$ is the Keplerian velocity at the surface of the protostar.
We cap the wind velocity at
100 km s$^{-1}$
in order to prevent it from becoming too severe a constraint on the time-step.
In addition, we apply the same limit to the \alfven velocity 
by injecting mass into cells with \alfven velocities exceeding this limit, as described in \citet{lee14}. This is particularly important in sink cells, and we verify that the total amount of mass injected during the course of the simulation is negligible.

\begin{figure}
\includegraphics[scale=0.42]{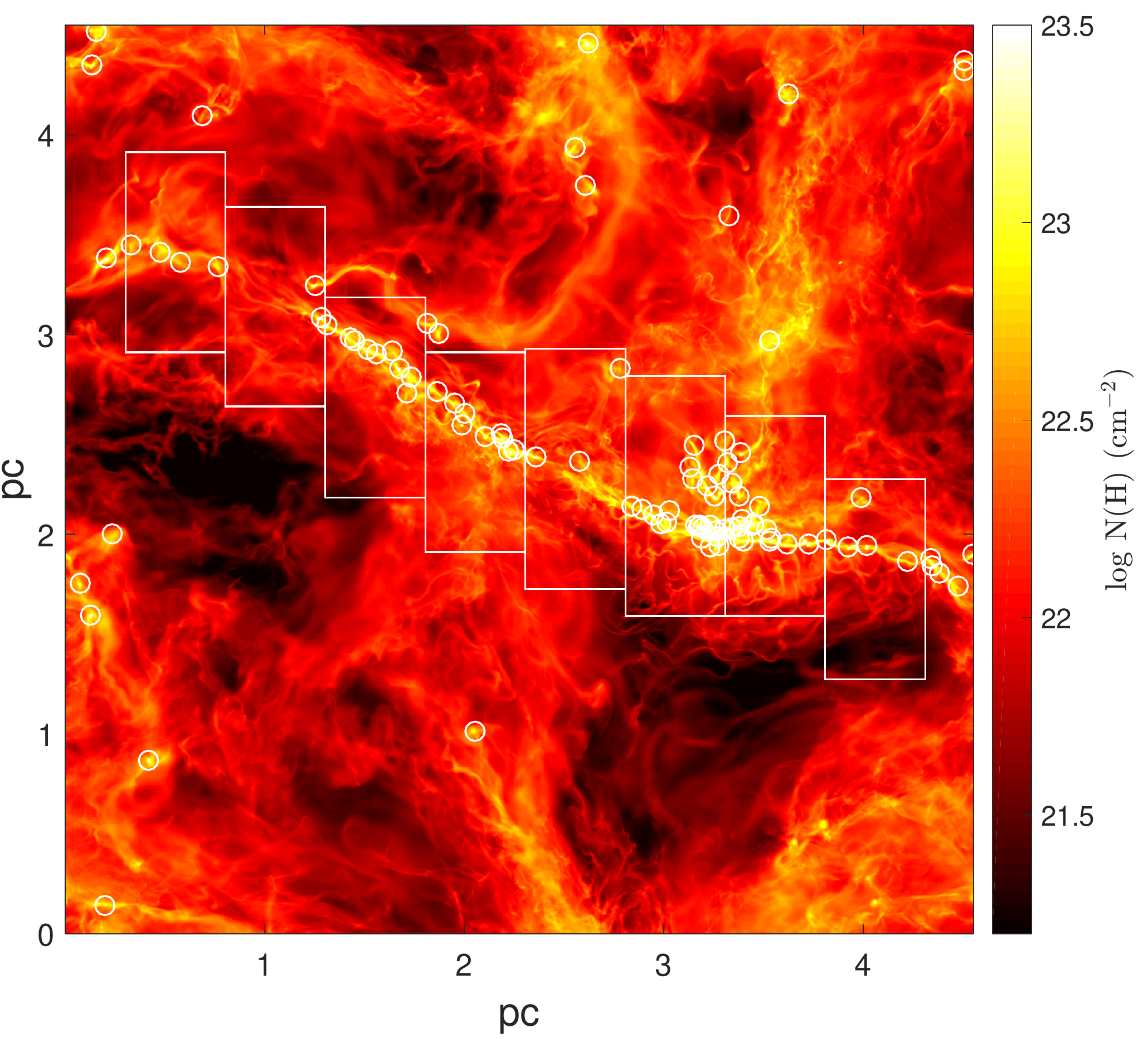}
\caption{Logarithmic-scale column-density map of the whole simulation region viewedd along the mean magnetic field direction at $0.5 \tff$. The rectangular boxes mark the zoom-in region in the simulation, in which AMR is allowed to refine up to a maximum of 6 levels. Protostars are allowed to form only inside the zoom-in region at the finest level. The total volume of the zoom-in region is 4.2 pc$^3$. The white circles mark the 100 most massive clumps identified at this time (see \citet{li15} on the discussion of 100 most massive clumps). 
\label{fig1}}
\end{figure}

To keep the computational cost down, we allow maximum refinement to occur only in a region with total volume $4.2 \,{\rm pc}^3$ surrounding the most massive long dark cloud filament as shown in Fig. \ref{fig1}.
The zoom-in region is composed of 8 rectangular sub-regions generally of size $\Delta x \times \Delta y \times \Delta z = 0.5 \times 1 \times 1$ pc.  Two of these sub-regions have size of $0.5 \times 1.2 \times 1$ pc in the region where the main filament collides with another massive filament cloud at a later time.
Inside this ``zoom-in'' region, there will be refinement up to a maximum 6 levels. Outside this region, there is no refinement and the resolution is at the base level of $\sim 1830$ au. In addition, inside the zoom-in region, the first 2 levels of refinement use the same criteria as in the IRDC simulation for refining pressure jumps, shear flows, and self-gravitating objects (the Jeans condition). From the third level, we refine only with the Jeans condition in order to reduce the computational time.

\subsection{Initial Conditions}

Simulations of isothermal MHD turbulence are scale-free (McKee e al. 2010) and are defined by two dimensionless parameters, the three-dimensional (3D) thermal Mach number, $\calm=\vrms/\cs$ (where $c_s$ is the thermal sound speed and $\vrms$ is the mass-weighted rms velocity) and the \alfven Mach number, $\ma=\vrms/\va$ (where $\va=\brms/(4\pi\bar\rho)^{1/2}$ is the mass-weighted \alfven velocity and $\brms=(\avg{B^2})^{1/2}$ is the square root of the volume-weighted mean-square magnetic field).

We need three dimensional relations \citep{mck10} to set the scale.
Interstellar molecular gas is observed to have a temperature $T=10T_1$~K with $T_1\simeq 1$, 
corresponding to an isothermal sound speed $\cs=0.188T_1^{1/2}$~km~s\e; this gives one relation. 
Most interstellar gas obeys the turbulent linewidth-size relation \citep{mck07},
\beq
\snt=0.72\spcs R_{\rm pc}^{1/2}=0.51\spcs\ell_{0,\,\pc}^{1/2}~~~\mbox{km s\e},
\label{eq:lws}
\eeq
where $\snt$ is the mass-weighted 1D nonthermal velocity dispersion in a sphere of radius $R$,
which we take to be the same as in a box of size $\ell_0=2R$, and where
$\spcs \sim 1$ allows for deviations from the typical linewidth-size relation.  
When self-gravity is included, a third dimensional parameter enters, the gravitational constant, $G$,
together with a corresponding dimensionless parameter that measures the effect of self-gravity, the
virial parameter, 
\beq
\avir\equiv \frac{5\sid^2 \ell}{2GM}\simeq \frac{5\snt^2 \ell}{2GM},
\label{eq:avir}
\eeq
where $\sid$ is the 1D mass-weighted velocity dispersion including the thermal velocity and where we have assumed that the flow is highly supersonic.

For highly supersonic, fully molecular gas, the physical parameters of the turbulent system---the size of the turbulent box, $\ell_0$,  
the flow time (or crossing time), $t_f$, the mass of the box, $M_0$, and the column density,
$\NH=\nbh\ell_0$---can
then be expressed as \citep[see the Appendix in][]{mck10}:
\begin{eqnarray}
\ell_0&=& \frac 23\;\frac{\calm^2 \cs^2}{(\spc^2/\mbox{ 1pc})}=0.0455\left(\frac{\calm^2 T_1}{\spcs^2}\right)~~\mbox{pc},
\label{eq:lscale}\\
t_f&=&\frac{\ell_0}{\vrms}=2.36\times 10^5\left(\frac{\calm T_1^{1/2}}{\spcs^2}\right)~~~\mbox{yr},
\label{eq:tscale}\\
\nbh&=&9.6\times 10^4\left(\frac{\spcs^4}{\avir\calm^2T_1}\right)~~~\mbox{cm\eee},
\label{eq:nscale}\\
M_0&=& 0.311\left(\frac{\calm^4T_1^2}{\avir\spcs^2}\right)~~M_\odot,
\label{eq:mscale}\\
\NH&=&1.34\times 10^{22}\left(\frac{\spcs^2}{\avir}\right)~~~\mbox{cm\ee}.
\label{eq:Nscale}
\end{eqnarray} 
The initial value of the uniform magnetic field is given by
\beq
B_0=4.56\;\left(\frac{\nbht T_1}{\beta_0}\right)^{1/2}\mmug
=31.6\left(\frac{\spcs^2}{\avir^{1/2}\ma}\right)\mmug.
\label{eq:bscale}
\eeq
where $\nbh$ is the average density of hydrogen nuclei, $\nbht=\nbh/(1000$~cm\eee), and $\beta_0=8\pi\rho \cs^2/B_0^2$ is the initial plasma beta.

The large-scale IRDC simulation has $\calm=10$, $T=10$~K, and $\avir = \spcs=1$, so that the size of the simulated turbulent region is $\ell_0 = 4.55$ pc, the number density $\nbh = 960$ cm$^{-3}$, the total mass $M_0 = 3110 \,M_{\sun}$, and the mean column density of the system $\NH = 1.34\times 10^{22}$ cm$^{-2}$. The free-fall time of the entire turbulent region based on the mean density,
\beq
\tff=\left(\frac{3\pi}{32G\bar\rho}\right)^{1/2}=1.37\times 10^6\nbht^{-1/2}~~~\mbox{yr},
\label{eq:tff}
\eeq
is $1.4\times10^6$ yr, which is $0.59 t_{\rm f}$. The magnetic field strength in the simulation is moderately strong, 
with $\mao=1$, corresponding to an initial plasma $\beta_0 = 0.02$. 

The relative importance of gravity and magnetic fields is measured by the normalized mass-to-flux ratio, $\mu_\Phi=M/M_\Phi$, where the magnetic critical mass is given in terms of the magnetic flux threading the cloud, $\Phi$, by $M_\Phi=\Phi/(2\pi G^{1/2})$. Gravity can overcome the field for $\mu_\Phi>1$, whereas gravitational collapse is impossible for $\mu_\Phi<1$. For a cubical box, $\mu_\Phi$ is related to the other dimensionless parameters of the problem by $\mu_\Phi=(5\pi/6\avir)^{1/2}\mao=1.62$,
which is only slightly supercritical. We have studied the magnetic field properties of the clumps in this simulation and reported the results in \citet{li15}.  We performed an additional initially weak mean magnetic field simulation 
with $\mu_\Phi=16.2$
in that study for comparison. By comparing the observations \citep{li09,li11} on the differences between magnetic field orientation of molecular clouds and the global mean field with our two models with different initial mean fields, we found that the moderately strong field model is necessary to explain the observed magnetic field orientation distribution.  Therefore, this work will focus on the star formation inside an IRDC with a moderately strong magnetic field.

\begin{table}
\caption{Initial Physical Properties of the Simulated Region}
\label{tab:region}
\begin{tabular}{lc}
\vspace{-0.3cm}\\
\hline
\hline
Entire simulation region \\
\hline
Thermal Mach number  & 10      \\
\alfven Mach number & 1         \\
Temperature   & 10 K            \\
Total mass    & 3110 $M_{\odot}$\\
Mean density  & 960 cm$^{-3}$   \\
Size          & 4.55 pc         \\
Virial parameter & 1            \\
Flow time     & $2.36\times10^6$ yr \\
Free-fall time& $1.4\times10^6$ yr \\
\hline
Zoom-in Region \\
\hline
Total mass    & 3110 $M_{\odot}$\\
Total volume  & 4.2 pc$^3$ \\
Mean density  & $3.80\times10^3$ cm$^{-3}$ \\
Mass-weighted median density & $1.47\times10^4$ cm$^{-3}$ \\
Free-fall time\fnm[1]& 212,500 yr \\
\hline
\end{tabular}
\begin{flushleft}
\fnt{1} {$^1$ Free-fall time based on the average density of gas above the mass-weighted median (see eq.\ref{eq:tffo}).}\\
\end{flushleft}
\end{table}

\begin{figure*}
\includegraphics[scale=0.9]{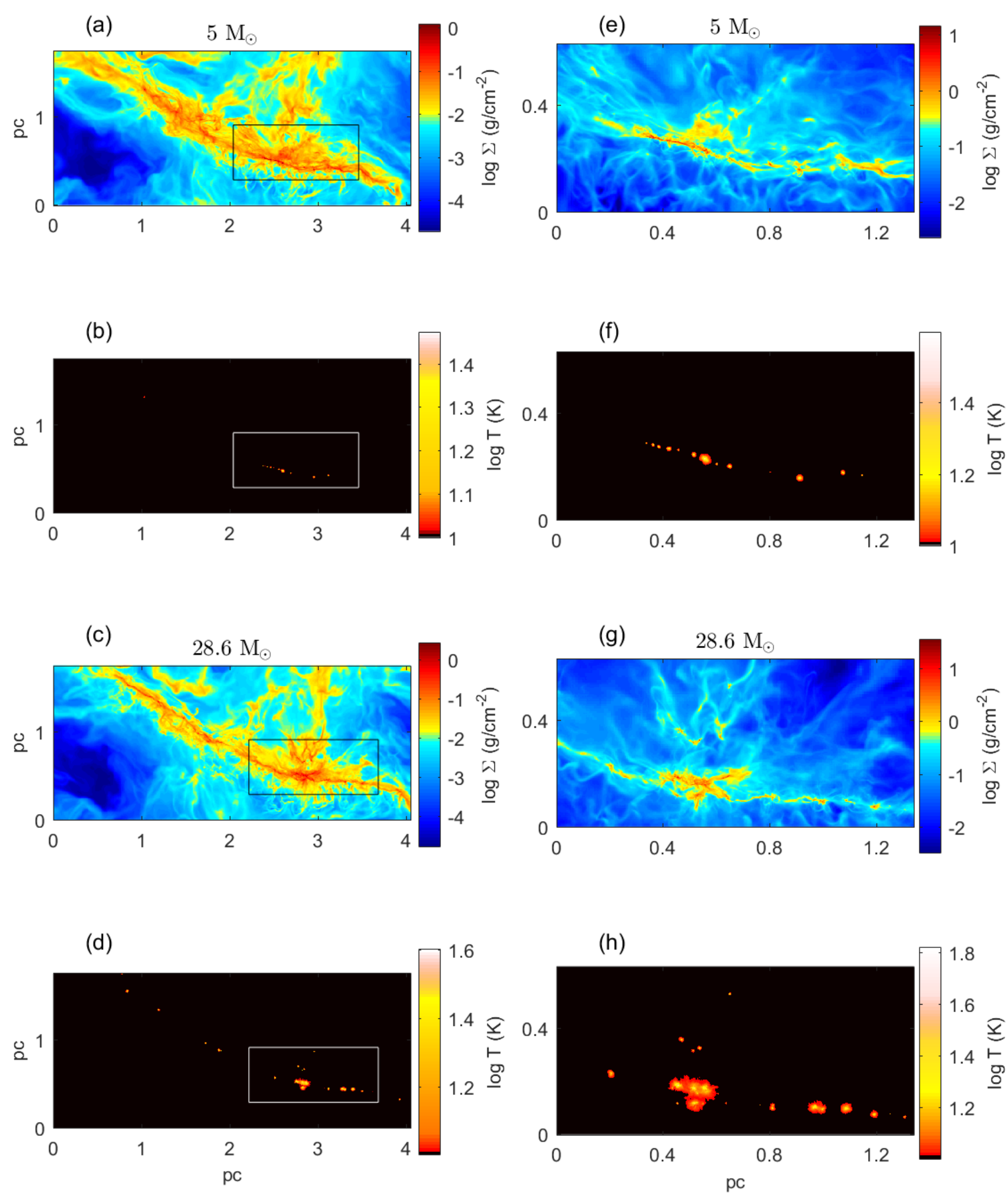}
\caption{(a) Logarithmic-scale column-density map
showing a 4 pc $\times$ 1.75 pc region around the main filament cloud
along the mean field direction at the time when the total mass of protostellar particles is 5 $M_\odot$. (b) Logarithmic-scale density-weighted temperature map at the time same as in (a). (c) Same as (a) but at the end of the simulation, when the total mass of protostellar particles is 28.6 $M_\odot$. (d) Same as (b) but at the time of (c). (e) - (f) an enlarged region as marked by the black (or white) rectangular box in the corresponding maps on the left. Complex filamentary structures are seen in the column-density maps. Initially, protostellar particles form mainly along the dark cloud filament, but 
at later times a clusters form at the junction between the two dark cloud filaments.
\label{fig2}}
\end{figure*}

In order to study the star formation in a moderately strong magnetic field and highly supersonic environment, we restart the simulation at a time of about $0.25 \tff = 350,000$ yr, when the highest density in the entire domain has not yet violated the Truelove condition \citep{tru97}. At this time, we increase the number of refinement levels to 6, corresponding to a maximum resolution of $\Delta x \simeq 28$ au.
The physical properties of the entire simulation box and the zoom-in region are summarized in Table \ref{tab:region}.
The \alfven\ Mach number is evaluated in the initial state, when the field is uniform, whereas the mass-weighted median density, $\rho_{\rm med}$, is evaluated at the end of the driving phase; the rest of the quantities apply at both times.

For convenience, in the rest of this paper, we shall set the time to be zero at this point when the zoom-in simulation is started.
The mean and mass-weighted median densities of gas inside the zoom-in region at this time are $3.80 \times10^3$ and $1.47\times10^4$ cm$^{-3}$, respectively,
significantly greater than the mean density in the entire simulation box, 960~cm\eee.
We define the characteristic free-fall time of gas in the zoom-in region by using the average density of the gas that is above the mass-weighted median density,
 $\bar\rho(\rho>\rho_{\rm med}) = 4.20 \times10^4$ cm$^{-3}$, which is
\beq
\tffo\equiv\left(\frac{3\pi}{32G\bar\rho(\rho>\rho_{\rm med})}\right)^{1/2}\simeq212,500 ~{\rm yr}
\label{eq:tffo}
\eeq 
at the beginning of the simulation.
We run the zoom-in simulation up to about 1.65 $\tffo$ (corresponding to a duration of 0.25 $\tff$).
At the end of the simulation, within the 0.3 pc width of the main filament, the overall mean $\muphi \sim 2.7$ and the mass per unit length of the filament is $\sim 5.8$ times the critical value \citep{inu97}.
The total computational cost of this zoom-in simulation is 4.8 million CPU hours with the simulation running parallel on 1024 processors.

\section{Simulation Results and Discussion}
\label{sec:results}

\subsection{Temperature, density and velocity structure in the zoom-in region}
\label{sec:temp}

\begin{figure*}
\includegraphics[scale=0.68]{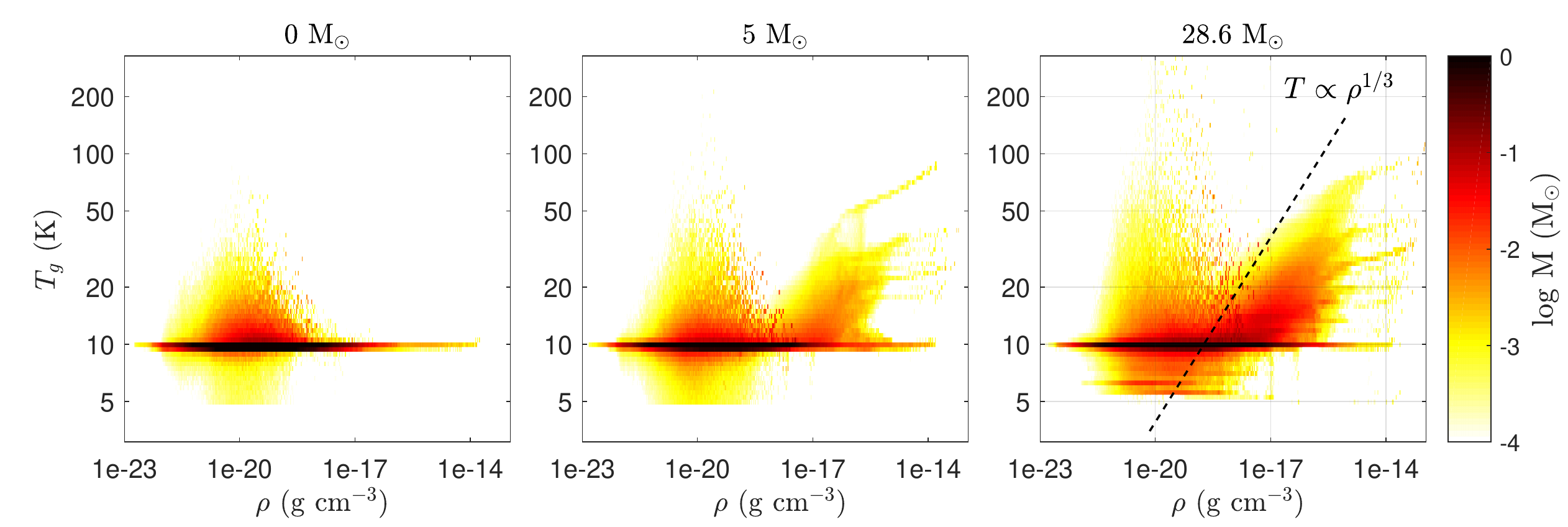}
\caption{$T_g - \rho$ relations of all cells in the simulation at time right before the first protostellar particle forms and when the total masses of protostellar particles are 5 and 28.6 $M_\odot$
(at 0.69 $\tffo$ and end of the simulation, respectively).
The bin sizes of density and temperature are both 0.025 dex in the diagram.
The heating of the gas is due primarily to the accretion luminosity of the protostars.
\label{fig3}}
\end{figure*}

\begin{figure*}
\includegraphics[scale=0.75]{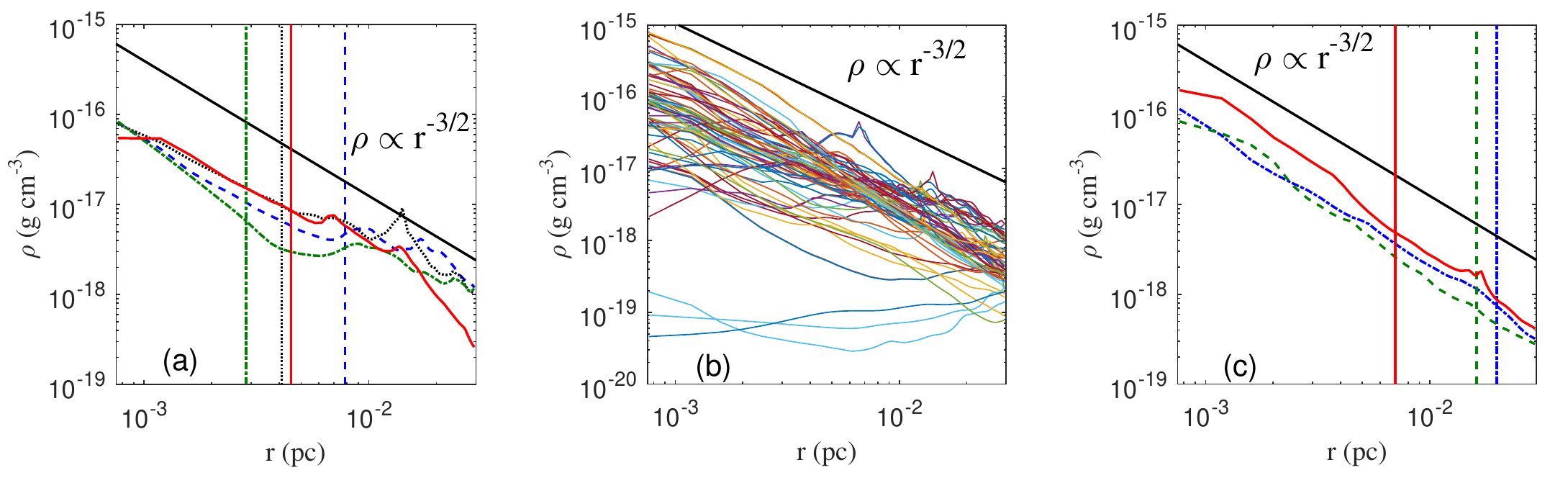}
\caption{(a) Four examples of the density profile of gas around single and multiple protostars with the corresponding radius of influence indicated by the vertical lines of the same line type at the end of the simulation. The inclined black line $\rho \propto r^{-3/2}$ is plotted for comparison. (b) All density profiles of gas around single and multiple protostars at the end of the simulation. (c) Three density profiles of gas around the same single star at an age of 60000 ($0.28 \tffo$, solid red), 120000 ($0.56 \tffo$, dashed green), and 315000 ($1.48 \tffo$, dot-dash blue) yr with the corresponding radii of influence indicated by the vertical lines of the same line type.
\label{fig4}}
\end{figure*}

Fig. \ref{fig2} portrays the column density, $\Sigma$, and the density-weighted mean gas temperature, $T_\Sigma$,
in the vicinity of the main filamentary cloud, where
\begin{eqnarray}
\Sigma &\equiv& \int_0^{\ell_{0z}} \rho dz, \\
T_\Sigma &\equiv& \frac{1}{\Sigma}\int_0^{\ell_{0z}} \rho T_g dz = \frac{\mu m_{\rm H} \ell_{0z}}{k_B \Sigma}\bar{P},
\end{eqnarray}
$\bar{P}$ is the mean pressure, $\mu=2.33$ is the mass per particle in units of the H mass, $m_{\rm H}$,
and $\ell_{0,z}=4.55$~pc is the size of the turbulent box.
These quantities are shown at two different times, one when the total mass in protostars is $5 M_{\odot}$ at 0.69 $\tffo$ and one at the end of the simulation when a total of $28.6 M_{\odot}$ is in protostars. The long filamentary
structure is obvious in the column density maps (Figs. \ref{fig2}a,c). Figs. \ref{fig2}e-h show the enlarged area of the junction where the vertical filament seen in Figs. \ref{fig2}a,c collides with the long horizontal filament. Many fine filamentary substructures are visible within the dark cloud filaments in the column density maps, similar to the dark filamentary dark cloud observations by {\it Herschel} \citep[e.g.][]{and10,kon15}.
These substructures are analyzed in Li, P.S. et al. (2017, in preparation).

As portrayed in Figs. \ref{fig2}b,f, a string of hot spots shows the protostars
are aligned in a 0.1 pc wide filamentary structure
that formed along the filament;
inspection of the filament in Fig. \ref{fig1} reveals the presence of dense cores, the progenitors of protostars. 
These results are similar to
the warm cores observed in the dark cloud G28.34+0.06, which is
at an early stage of star formation \citep{zha15}.
Figs. \ref{fig2}b, d, f, and h show the corresponding density-weighted temperature maps. We note that by the end of the simulation, the number of protostars per unit length (51 protostars in a 1 pc region) 
at the junction of the colliding filaments is about 6 times greater than in the rest of the filament.
The ratio of protostellar mass to gas mass in the colliding region is 0.067, whereas it is only 0.028 in the rest of the filament.
Many protostars in the colliding region are relatively young, and therefore the average stellar mass there is only about
1/3 of that in the rest of filament.
This shows how cloud-cloud collisions can enhance star formation.

Fig. \ref{fig3} shows the density-temperature phase relation of all the cells in the simulations at the time
 right before the first protostar forms, and at times when the total masses of protostars are 5 and 28.6 $M_\odot$.
The gas mass inside the zoom-in region at these three times are 555.9, 607.4, and 636.3 $M_\odot$.
The increase in the total mass inside the zoom-in region indicates that the star-forming region continues to slowly accrete mass.
Before formation of the protostars, 99 per cent of the mass is at $10\pm1$ K. A small fraction of the mass has a temperature outside this range due to adiabatic heating from supersonic shocks and cooling from rarefactions.
At late times some of the high-density gas reaches temperatures near 100 K due to heating by the accretion luminosity of the protostars; heating due to the intrinsic stellar luminosity is small since most of the protostars are less than a solar mass.
There is a tiny fraction of gas at even higher temperatures, $T> 1000$ K, associated with the shocked protostellar outflow, but this is not shown.

A striking feature in Fig. \ref{fig3} is the development of a triangular shape distribution of $T_g$ vs $\rho$; such a feature also appears in previous simulations of cluster formation with radiative transfer \citep[e.g.][]{kru11,mye14}. 
This feature is associated with gas near the protostars, so we first need to determine the density profile of gas near a
protostar. In
Fig. \ref{fig4}, we compute the density profiles around 4 protostars at the end of the simulation,
using the procedure described in \citet{mur17}: We average over concentric spherical shells of thickness 3 cells starting from the edge of the accretion zone, which has a radius of 4 cells. \citet{mur17} defined the ``radius of influence" as the radius at which the mass of enclosed gas equals the mass of the central star and any protostellar disk, and they argued that the density scaled as $r^{-3/2}$ inside this radius. Our results are consistent with this, even though
these four examples include single and multiple protostellar systems. Fig. \ref{fig4}b shows profiles of the density distribution around all the protostars at the end of the simulation.
About 85 per cent of the protostars show $r^{-3/2}$ profiles that extend to the radius of influence, and of these about
70 per cent 
have an $r^{-3/2}$ profile extending out to at least twice of the radius of influence.
About 15 per cent of the protostars have shallower or even dipping profiles near their centers. Some of these anomalous profiles are the result of outflows inside multiple systems that create lower density voids, some of them have consumed most of the gas in the vicinity,
and some of them have very dense gas streams nearby that create irregular density profiles.
In Fig. \ref{fig4}c, we plot the density profiles and the corresponding radii of influence of a single star 
at three different epochs, $t=6\times 10^4, 1.2\times10^5,$ and $3.15\times 10^5$ yr. The profiles maintain a slope of about -3/2 but shift up and down somewhat in time. On average, the density profile is roughly independent of time provided there is not a deficit of gas near the protostar. For the several protostars that do not have much gas around them and experience a diminishing accretion rate, the density profiles have much shallower slopes.

\begin{figure}
\includegraphics[scale=0.45]{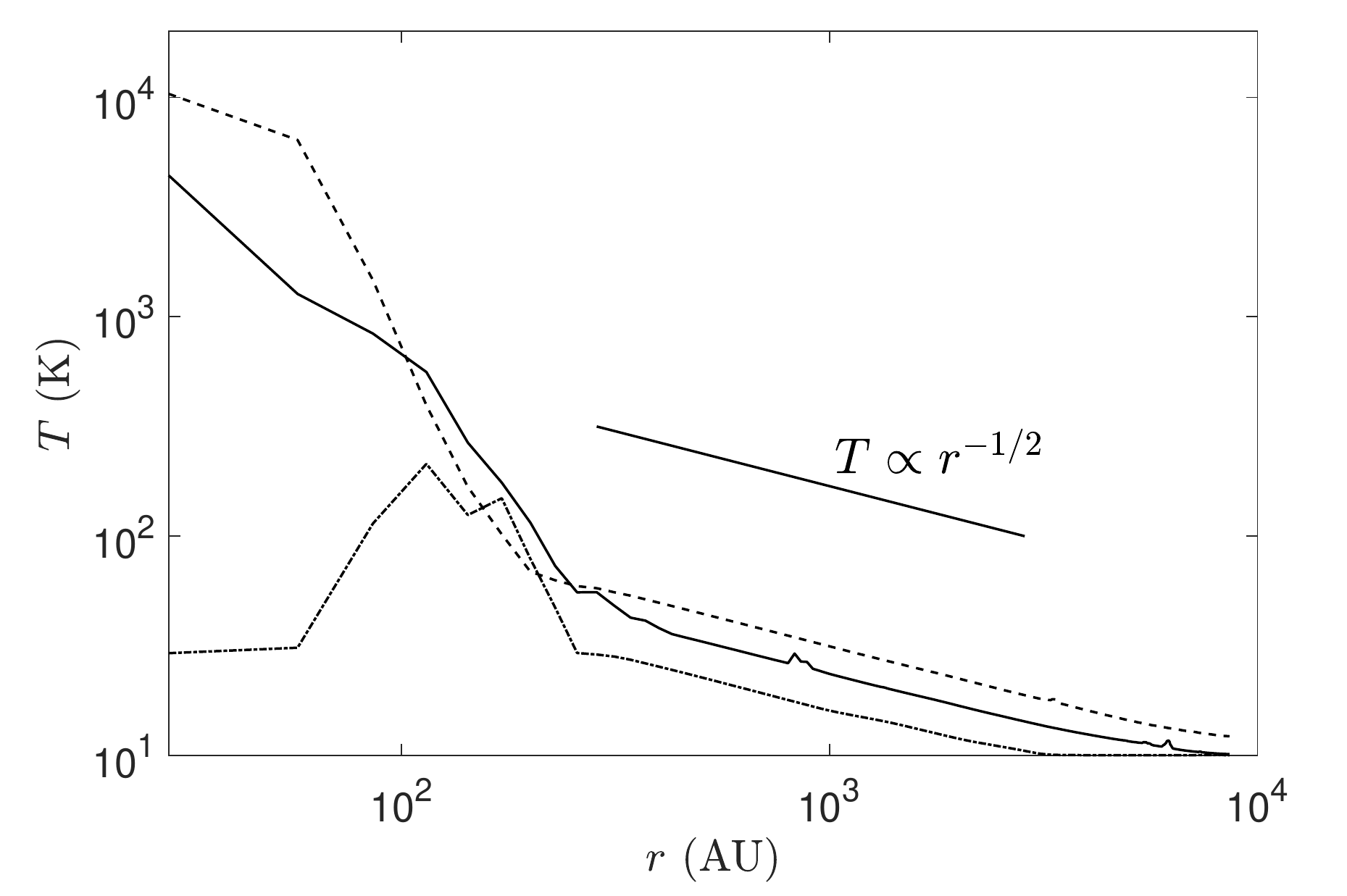}
\caption{Temperature profile of three typical regions around protostars.  Except the inner part ($\le 300$ au), the temperature profile shows a typical $T \propto r^{-1/2}$ profile of an optically thin region heated up by a blackbody radiation source 
\label{fTr}}
\end{figure}

\begin{figure}
\includegraphics[scale=0.35]{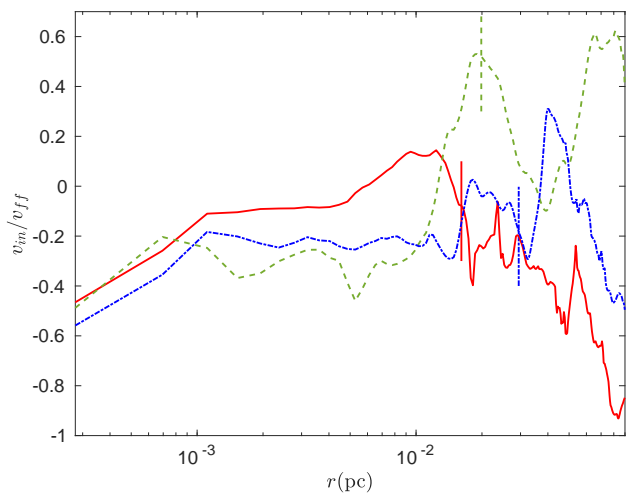}
\caption{Three examples of the radial velocity of gas around protostars, normalized by the freefall velocity. The short, thick vertical lines indicate the corresponding radii of influence. 
\label{frvel}}
\end{figure}

\begin{figure*}
\includegraphics[scale=0.75]{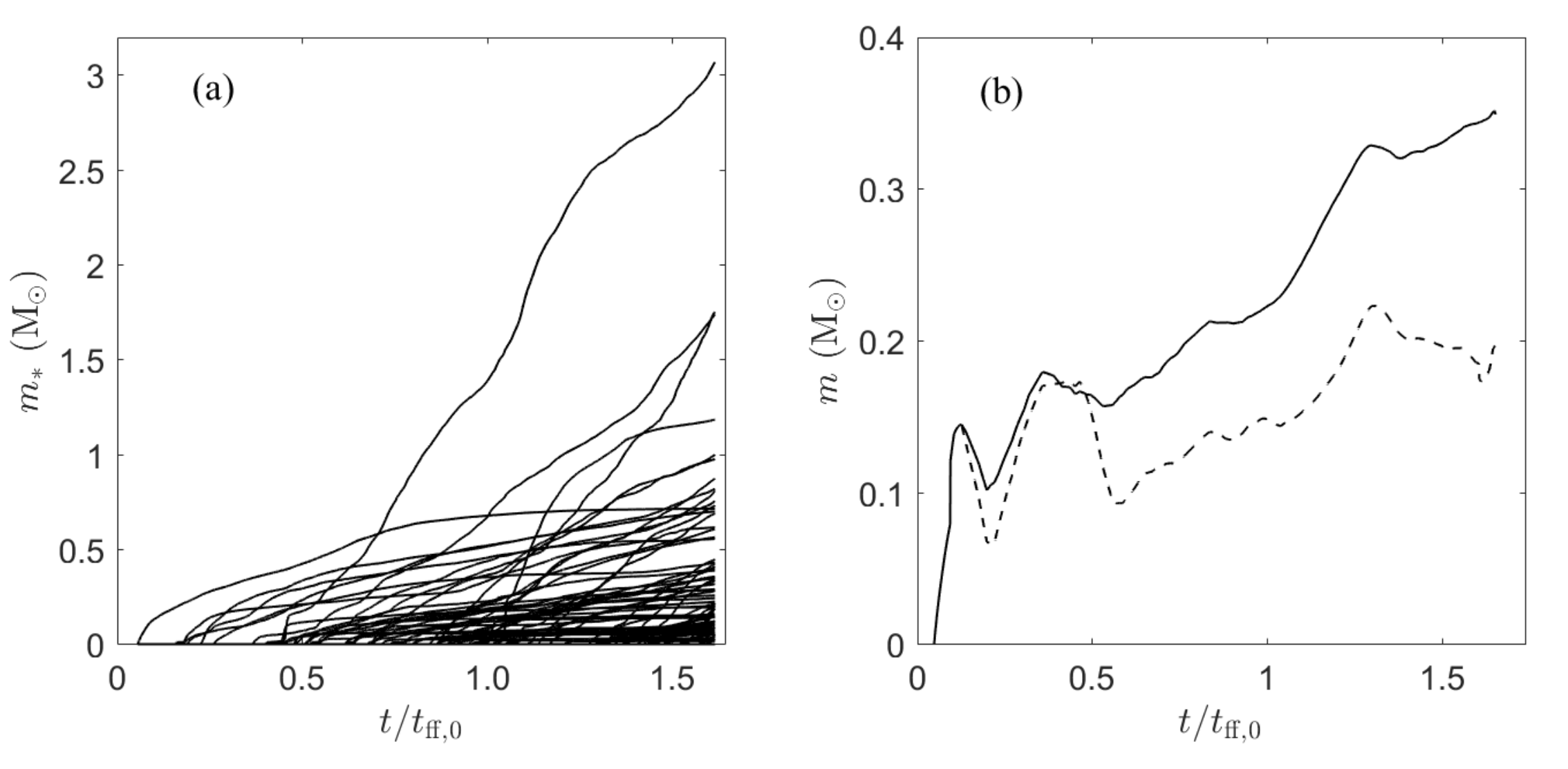}
\caption{(a) Accretion history of the protostellar particles. (b) The time evolution of the mean (solid curve) and median (dashed curve) protostellar mass. The two curves are smoothed over $2\times10^4$ yr. The time is measured in units of $\tffo$.
\label{fig5}}
\end{figure*}

\begin{figure*}
\includegraphics[scale=1]{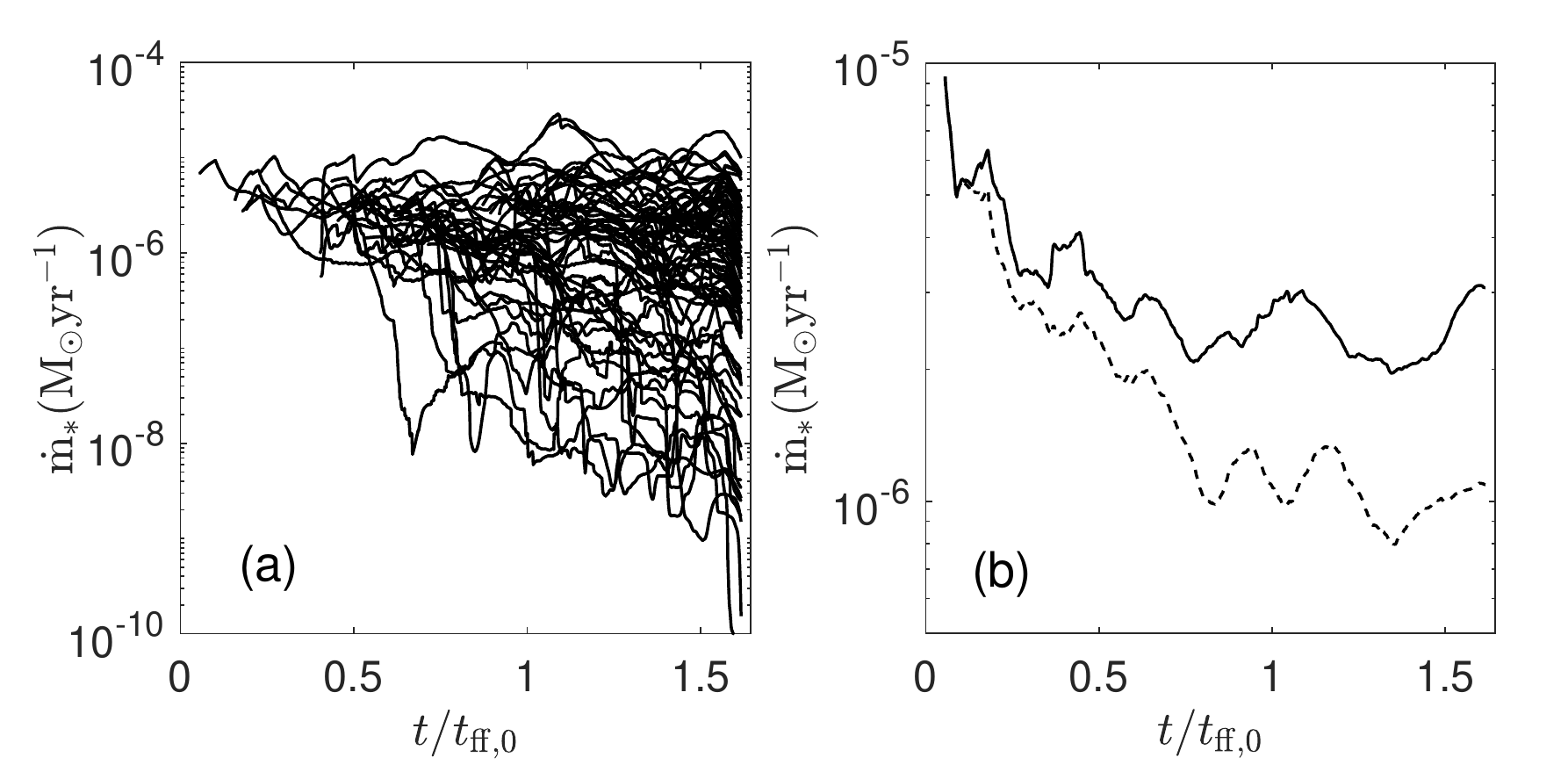}
\caption{(a) Mass accretion rate of the protostellar particles. Most of the particles accrete at $\sim 10^{-5} - 10^{-7} M_\odot {\rm yr}^{-1}$. (b) Time evolution of the mean (solid) and median (dash) values of the mass accretion rate per protostar. All the curves are smoothed over $2\times10^4$ years. The time is measured in units of $\tffo$.
\label{fig6}}
\end{figure*}

Having determined the density profile around the protostars, we can now infer the temperature profile.
In \textsc{orion2}, we make the approximation that the emission and absorption opacities are the same, which is valid if the dust temperature is close to the radiation temperature. We then have
$L/r^2 \propto T^4$ and $T\propto r^{-1/2}$, as shown in Fig. \ref{fTr}.
As noted above, the density profile within about 300 au of the protostars in most cases is approximately given by $\rho \propto r^{-3/2}$. The density covered by the triangular pattern is from $\sim 10^{-18} - 10^{-15}$ g cm$^{-3}$, which is about the same range as near the protostars (Fig. \ref{fig4}). Combining this with the $T\propto r^{-1/2}$ temperature profile, we find $T \propto \rho^{1/3}$ in the optical thin limit. One can see that this is in good agreement with the upper envelope of the points in the $T-\rho$ plane. The fan shape is the result of all the contributions from the gas heated by stars of different luminosities and of the density variations at a given distance from individual stars. To verify that the triangular pattern is due to protostellar heating, we found that it disappeared when we removed the cells around the protostellar particles. Most of the gas around the protostars has a temperature $\ga 50$~K within 250 au and $\ga 20$~K within 1000 au, depending on the accretion luminosity.

\citet{mur17} found that the radial infall velocity of gas is about 25 per cent of the freefall velocity inside the radius of influence.
For the stars in our simulation, over the range 200 au from the star to the radius of influence,
we find that the mean radial velocity of the gas is about 30 per cent of the freefall velocity, 
similar to their results. In Fig. \ref{frvel}, 
we plot three examples of the radial velocity, normalized by the freefall velocity of gas. The infall velocity remains a small fraction of the freefall velocity 
for $r>10^{-3}$~pc $\simeq 200$~au. The increase of the magnitude of this ratio inside that radius could be due to numerical effects, since the artificial accretion zone has a radius of 112 au.

\subsection{Accelerating star formation}
\label{sec:sfr}
\begin{figure*}
\includegraphics[scale=0.75]{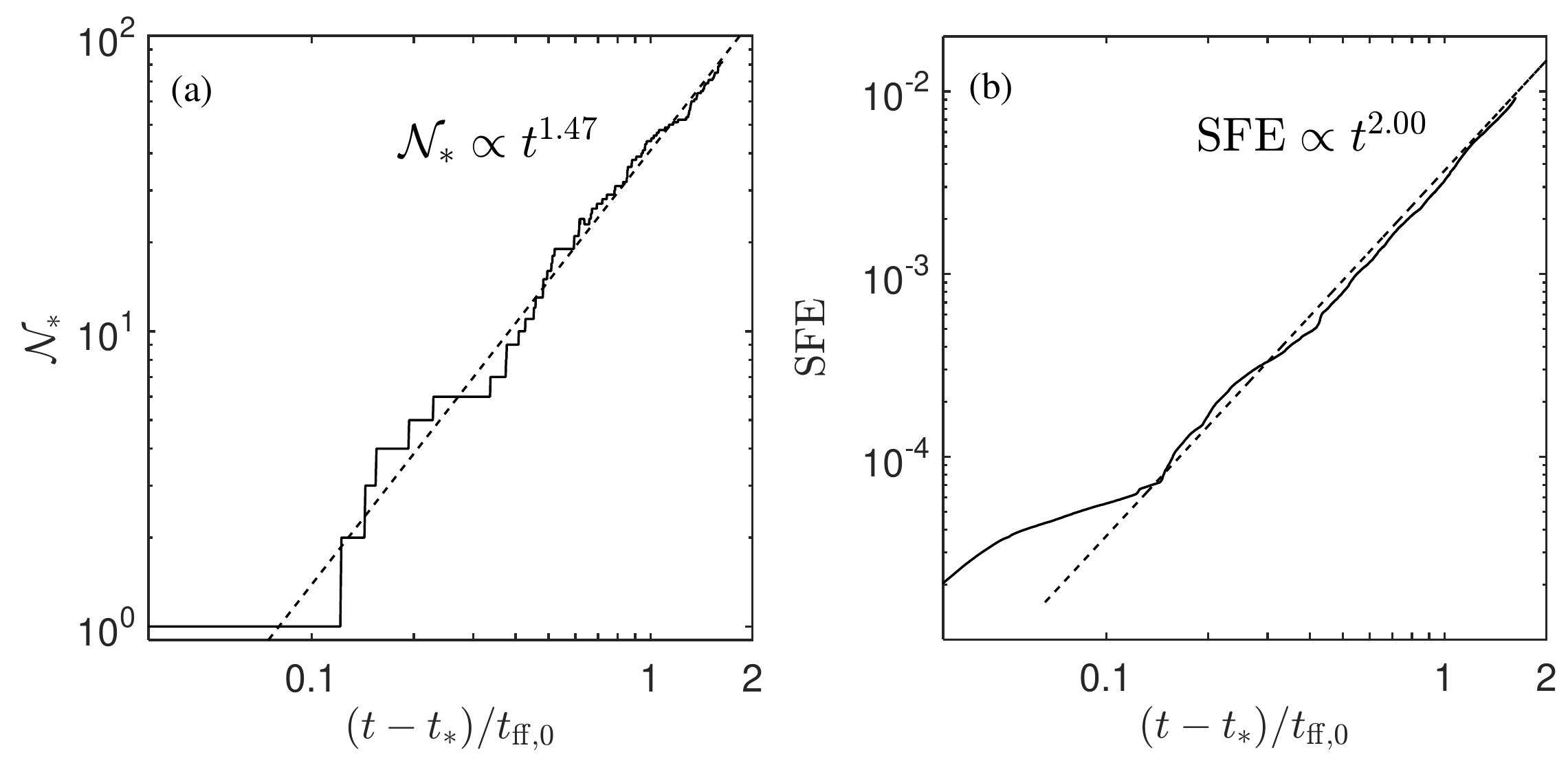}
\caption{(a) Number formation rate of protostellar particles, (b) the power law relation of star formation efficiency and time. $t_*$ is the time when the first star particle is formed.
\label{fig7}}
\end{figure*}

\subsubsection{Accretion histories}

There are a total of 82 protostellar particles created in the simulation. In Fig. \ref{fig5}a, the history of mass growth of all the protostars is shown. The most massive protostar has $3.07 M_\odot$ and is continuing to accrete mass at the end of the simulation, as seen from the slope of the accretion curve.
In Fig. \ref{fig5}b, we plot the time evolution of the mean and median protostellar mass.  The median mass fluctuates around $0.2 M_\odot$  after $1.2 \tffo$ and the mean value increases slowly.
In Fig. \ref{fig6}a, we plot the mass accretion rate smoothed over
$2\times10^4$ years for easy visual inspection. Most of the protostars are accreting at rates of $10^{-5} - 10^{-7} M_{\odot}$, as expected from observations.
The accretion rates vary and a few protostars have accretion rates below
$10^{-8} M_{\odot} {\rm yr}^{-1}$.
Some of these protostars have a quick drop in accretion rate of more than 2 orders of magnitude to $\la 10^{-8}\, M_{\odot} \,{\rm yr}^{-1}$ and remain low to the end of the simulation. The gas around these few protostars has a fairly flat, low-density profile, as shown in Fig. \ref{fig4}b and discussed above. Some show a somewhat steady drop in the accretion rate and reach $\sim 10^{-8}\, M_{\odot} \,{\rm yr}^{-1}$ at the end of the simulation.  Finally, others have fluctuating accretion rates that drop to $\sim 10^{-8} \,M_{\odot}\, {\rm yr}^{-1}$ and then rise back up to $\sim 10^{-7}\, M_{\odot}\, {\rm yr}^{-1}$.
However, all the protostellar particles have at least $0.4\,M_\odot$ of gas within 0.05 pc and thus approximately satisfy
the criterion imposed by \citet{dun13} that a protostar must have a detectable amount of gas around it (they give a range of
$0.5-0.8\,M_\odot$ for the minimum detectable mass 
for the stars in their sample).\footnote{At the end of the simulation, the total mass of gas within 0.05 pc of the 82 star particles is $94.5\,M_\odot$, a significant fraction of the $636.3\,M_\odot$ of gas in the zoom-in region. As a result, the mean density of the gas near the protostars, $2.2\times 10^4$~cm\eee, is comparable to the mean
density of gas above the mass-weighted median density in the zoom-in region, $4.20\times 10^4$ cm\eee.}
There are four protostars that have accretion rates $\ga 10^{-5}\, M_{\odot} \,{\rm yr}^{-1}$ 
for at least some short period of time. One of them has an accretion rate at or above this value from its creation to the end of the simulation 260,000 years later and becomes the most massive protostar in the simulation. 

For each protostar, we have determined its mean accretion rate after it reaches a mass of $0.04\,M_\odot$;
we set that threshold since we do not accurately treat the formation of the first core \citep{lar69}, which can have a mass
of about
this value in our simulation \citep{vay16} (see discussion in Section \ref{sec:pmf}).
The mean and median accretion rates of the protostars are $3.9 \times10^{-6}\,M_\odot$~yr\e\ and $1.4\times10^{-6}\,M_\odot$~yr\e, respectively.
The dispersion in the log of the mean accretion rates of the protostars is 0.54 dex.
We can also evaluate the dispersion of the accretion rate of each protostar; for example, a given protostar might have a mean accretion rate of $2.5\times 10^{-6}\,M_\odot$~yr with a one-sigma dispersion of 0.4 dex.
The average value of this dispersion in the log of the accretion rates of individual stars is 0.51 dex. 
Thus, the variation in accretion rates between different protostars is comparable to the variation in the accretion rates of individual protostars.
The time evolution of the mean and median values of the mass accretion rate of all the protostars is shown in Fig. \ref{fig6}b. The rates drop and reach asymptotic values at late times of about $2.5\times10^{-6}$ and $1\times10^{-6}M_{\odot}\, {\rm yr}^{-1}$ for the mean and median, respectively. 

\subsubsection{Results for the number and mass of stars produced}

Fig. \ref{fig7}a shows the increase in the number of protostars, $\caln_*$, with time. The time plotted is measured from the time when the first protostellar particle is formed, $t_*$. We find $\caln_*\propto t^{1.47}$, indicating an accelerating rate of star formation.  Such accelerating star formation has been seen in several star-forming clusters \citep[e.g.][]{pal00,ryg13}. 

To evaluate the mass of stars produced, we use the
star formation efficiency (SFE), which is defined as the total mass in protostars divided by the total mass of the cloud plus stars,
\beq
{\rm SFE} \equiv \frac{M_*}{M_g+M_*},
\label{eq:sfe}
\eeq
where $M_*$ is the total mass of protostellar particles
and $M_g$ is the total mass of gas in the region under consideration.
In this section we are interested in the time dependence of SFE rather than its value, so we shall evaluate this just for the 
zoom-in region.
The time evolution of the SFE is well fit by SFE $\propto t^2$ (see Fig. \ref{fig7}b), which implies that the cluster mass grows as $M_*\propto t^2$ and that the star formation rate grows as $\dot M_*\propto t$.

\subsubsection{Comparison with previous work}

A stellar mass that increases approximately quadratically in time has been 
suggested theoretically for the formation of individual stars by \citet{mck03} and \citet{mur15}; the latter authors suggested that their model also applies to cluster formation, but no details were given. A $t^2$ increase in the stellar mass, which is proportional to the SFE, for a cluster has been
found in previous simulations:
We estimate that the SFE in the simulations of \citet{pad14} varies approximately as $ t^{1.75}$, and the simulations by \citet{lee15} and \citet{mur17} have an SFE $\propto t^2$.
The simulation in \citet{pad14} is an isothermal MHD turbulence simulation using 6 levels of refinement with an initially magnetic field of $\ma = 5$ and is driven at Mach 10. The maximum resolution is 50 au, about half the resolution of our zoom-in simulaiton. The stars in the simulation are the sink particles created at the finest level. Only half of the accreted gas is given to the sink particles.  The other half is removed from the simulation without feedback. The simulations in \citet{lee15} are isothermal MHD turbulent simulations on a unigrid with an initially weak magnetic field and are driven at Mach 9. The grid size, $\sim 0.03$ pc, is $> 200$ times larger than that of our zoom-in simulation. The protostellar particles created in these simulations are very massive due to the low resolution. The simulation in \citet{mur17} is a pure hydrodynamic turbulent model using 8 levels of refinement to achieve resolution about 100 au, about four times worse than our zoom-in simulation. There is no radiative or outflow feedback in either of these simulations.
In contrast to these simulations, our simulation includes both radiative and outflow feedback, but these feedback effects do not significantly affect the power law in the temporal evolution of the SFE.

A factor that does affect the outcome of the simulations is whether the simulations are driven or not. The simulations in \citet{pad14} and \citet{lee15} are driven ideal MHD turbulent simulations, just as ours is. On the other hand, the simulations of \citet{mye14} have decaying turbulence, and in these simulations the SFE increases more rapidly with time: SFE$\propto t^{3.4}$ in the hydrodynamic case and $t^{2.7}$ in the MHD case with a moderately strong magnetic field. These results show that driven turbulence provides a significant support against gravitational collapse and reduces the SFE,
at least at late times.
Insofar as driven turbulence approximates the injection of turbulent energy from larger scales, this result suggests that turbulence reduces the SFE in nature as well.

\begin{figure*}
\includegraphics[scale=0.75]{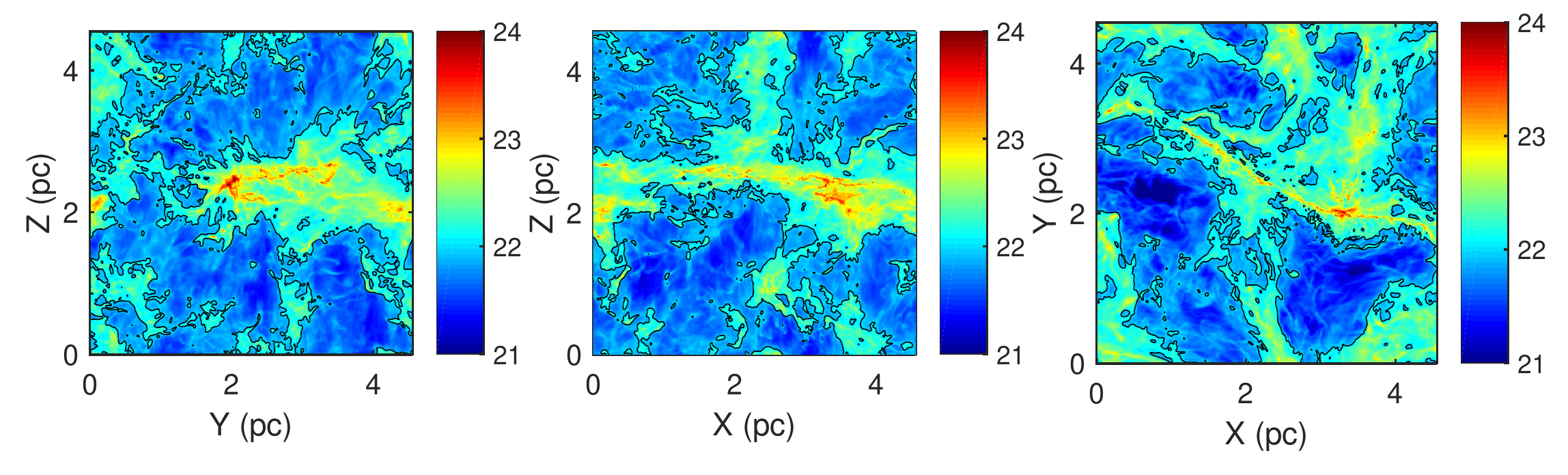}
\caption{Logarithmic scale column density maps along 3 cardinal axes with the contour at $1.03\times10^{22} {\rm cm}^{-2}$, corresponding to extinction $A_K = 0.8$ mag.
\label{fig8}}
\end{figure*}

\subsection{The star formation rate per free-fall time, $\eff$}
\label{sec:epsff}

To compare the star formation rate (SFR) in different star formation regions with different densities, 
the dimensionless SFR per free-fall time is introduced \citep{kru05},
\beq
\eff \equiv \frac{\dot{M}_{*,{\rm obs}}}{M_g/t_{\rm ff}},
\label{eq:epsilon}
\eeq
where
\beq
SFR\equiv\dot{M}_{*,{\rm obs}}=M_*/t_{\rm SF}
\eeq
is the average rate of star formation over the time interval over which the stars have been forming, $t_{\rm SF}$. Here, we measure $t_{\rm SF}$ from the time that a star first form to the end of the simulation.
The star-formation rate per free-fall time is related to the SFE (Equation \ref{eq:sfe}) by
\beq
SFE=\frac{\eff(\tsf/\tff)}{1+\eff(\tsf/\tff)}\simeq \eff(\tsf/\tff),
\eeq
where the second step follows for the usual case in which $\eff$ is small and $\tsf/\tff$ is not large.

\citet{kru12a} 
emphasized the importance of using a reasonably good estimate of the volume density of the star-forming gas in estimating $\eff$ from observations. From our simulation, we determine $\eff$ using a modification of the approach of \citet{kru12a}: Since star formation occurs in dense gas, we focus on gas in the upper half of the density distribution--i.e., above the mass-weighted median density, $\avg{\rho}_M$. They evaluated $\tff$ at $\rhomm$, but it is more accurate to evaluate it at $\bar\rho(\rho>\rhomm)$, the average density of gas denser than $\rhomm$. This theoretical value of $\eff$ is then
\beq
\efft \equiv \frac{\dot{M}_{*,{\rm obs}}}{0.5M_g/t_{\rm ff}(\bar\rho(\rho>\rhomm)},
\label{eq:epsilont}
\eeq
where the 0.5 comes from the fact that half the gas mass has $\rho>\rhomm$.
In simulations, the star formation time, $t_{\rm SF}$, is the duration from the time the first star is formed to the time of the SFR measurement. If the first star forms anomalously early, as in our simulation (see Fig \ref{fig7}), this time can be adjusted. Since the density of the star-forming gas evolves with time, the density is evaluated at the same time that the SFR is, just as an observer would do.
The value of $\bar\rho(\rho>\rho_{\rm med})$ inside the zoom-in region at the end of the simulation is $1.47\times10^5$ cm$^{-3}$. The corresponding $\tff = 1.13\times 10^5$~yr and since the gas mass $M_g$ inside the zoom-in region is $636\,M_{\odot}$, it follows that $\efft = 0.03$. At an earlier time, when $M_*=5 M_{\odot}$, we find $\efft = 0.018$, indicating an increase of $\efft$ in time. The major factor leading to the increase is that $\dot{M}_{*,{\rm obs}}$ increases approximately linearly with time.

A recent study of the star formation rate from Galactic clouds to high-redshift galaxies by \citet{kru12a,kru13} shows that there is an almost universal value of $\eff$ based on the volumetric star formation law. Their best fit of $\eff$ is 0.015 with a scatter of a factor of 2.7.
A comparable value of $\eff$ has been obtained in a study of Class I sources in dense molecular clumps in the Galaxy by \citet{hey16},
who find $\eff=0.02\pm0.01$ after correcting for the fact that they are insensitive to protostars with $m_*\la 1-2\,M_\odot$. The results of \citet{lee16} give a median consistent with this value of 0.01, but with a much larger dispersion of 0.86 dex.
The larger dispersion is due to their inclusion of clouds with high-mass star formation, which are subject to feedback effects; such star-forming regions were therefore excluded from the analysis of \citet{kru12a}.
In any case, our simulation does not include any massive stars (the maximum value of $m_*$ is $3\,M_\odot$),
so the results of \citet{lee16} are not relevant for our simulation.
The median of $\eff$ of the Milky Way clouds is tabulated by \citet{kru12a} is 0.017. 
Some of the Milky Way molecular clouds have values of $\eff$ up to 6 times greater than the median; they point out that this could be because these clouds are filamentary, whereas the density, and therefore $\tff$, was estimated under the assumption that they are spherical.   

\begin{figure*}
\includegraphics[scale=0.75]{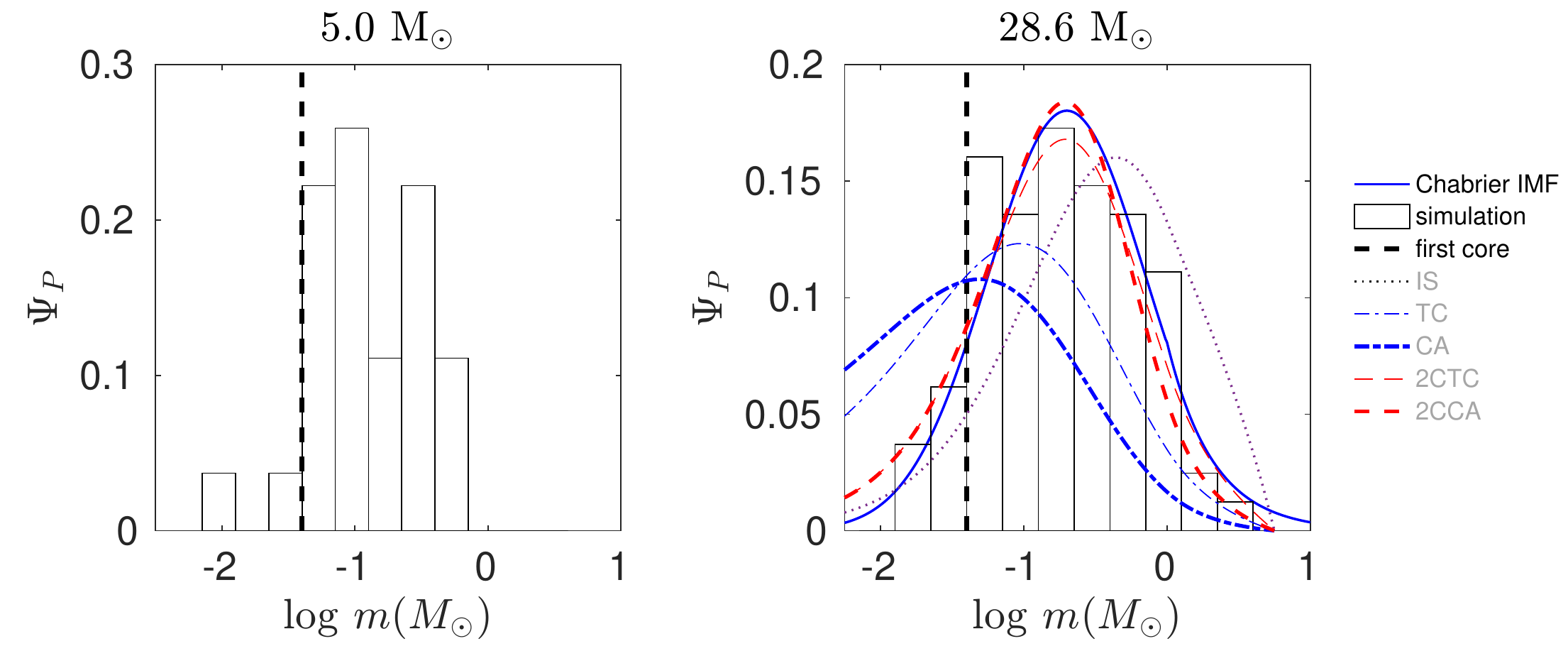}
\caption{(a) The PMF of the protostellar particles at the time when total mass of particles is $5 M_\odot$  The vertical thick dashed line indicates the first core mass of $0.04 M_{\odot}$ expected in our environment. The histogram to the left of this line is uncertain (see Section \ref{sec:pmf} for discussion). (b) The PMF of the protostellar particles at the end of the zoom-in simulation, overplotted with five theoretical PMFs: IS (black dotted), 2CCA (thick red dashed), 2CTC (thin red dashed), CA (thick blue dot-dashed), TC (thin blue dot-dashed), and Chabrier's IMF (blue solid). The cutoff mass of the five PMFs is 5.6 $M_{\odot}$. 
\label{fig9}}
\end{figure*}

\subsubsection{Simulation results for $\eff$}

The first step in evaluating $\eff$ is to determine $\tff$, which we do by using the extinction method described in \citet{kru12a}: First we estimate the area of the cloud with $\Sigma \ge 14.5 M_\odot {\rm pc}^{-2}$ for $A_K = 0.1$ mag, or $\Sigma \ge 115 M_\odot {\rm pc}^{-2}$ for $A_K = 0.8$ mag. In the case of $\Sigma \ge 14.5 M_\odot {\rm pc}^{-2}$, the area with a column density higher than this is basically the whole simulation region, so $\tff$ is determined by the mean density of the box. If we assume a spherical cloud with a projected area equal to the projected area of the box, the volume will be a factor of $4/(3\sqrt{\pi})=0.752$ times that of the box, so the mean density is $ 3.0 \times 10^{-21}$ g cm$^{-3}$. The value of $\tff$ estimated from this approach is $1.22 \times 10^6$ yr, so $\eff= 0.049$.

However, we need to make a correction for the fact that in our simulation, stars can form only in the zoom-in region: The sink particle algorithm allows protostellar particles to form only at the highest refinement level, and since there is no refinement outside the zoom-in region, no stars can form there. To estimate the value of the SFR that we would have obtained had we allowed star formation to occur throughout the box, we use the clump information from our recent work \citep{li15}, in which we studied the 100 most massive clumps in the simulation. At 0.5 $\tff$, corresponding to the end of the zoom-in simulation, there were 78 clumps located inside the zoom-in region (see Fig. \ref{fig1}). These clumps include 2/3 of the densest clumps in the entire simulation box, comparable to the fraction expected from a sample of 78 per cent of the clumps. 
If we assume that the proportion of stars formed in the zoom-in region is similar to the fraction of the most massive clumps there, then the total SFR in the simulation would be larger by a factor of 1/0.78. 
This increases $\eff$ by a factor 1/0.78 to $\eff \simeq 0.067$, which is at the high end of the values of $\eff$ determined with the extinction method by \citet{kru12a}.

Since the entire simulation region has a column density above $14.5 M_\odot {\rm pc}^{-2}$, it is better to use a higher extinction limit, $A_K = 0.8$ mag, to estimate the star formation rate in filamentary dark clouds in the simulation. The region with column density $\Sigma \ge 115 M_\odot {\rm pc}^{-2}$ (or $N(H) \ge 1.03\times10^{22} {\rm cm}^{-2}$) is smaller and more filamentary, as shown in Fig. \ref{fig8}. Using the above mentioned procedure, the mean column densities obtained from the three cardinal axis projections are between 6.88 and $8.70 \times 10^{-21}$ g cm$^{-3}$. The corresponding values of $\eff$ for the stars in the zoom-in region are between 0.027 and 0.028, closer than one might expect due to accidental cancellations. 
These values are
basically the same for the three projections and quite close to the theoretical value from Equation (\ref{eq:epsilont}).
All 100 of the most massive clumps are found to be inside the $A_K = 0.8$ mag contour.
Applying the correction of star formation suppression outside the zoom-in region, 
the corrected $\eff = 0.036$ for the whole region, about half that obtained using $A_K = 0.1$ mag.
This is close to the theoretical $\efft$ of 0.031 obtained above.
In \citet{kru12a}, there are seven galactic clouds that are in highly filamentary form. The median values of $\eff$ for these seven clouds are 0.0349 and 0.0144 for $A_K=0.1$ and 0.8. The corresponding values of $\eff$ from the simulation are about twice as large.

\subsection{Protostellar Mass Function and the Initial Mass Function}
\label{sec:pmf}

Here we construct the mass function of the protostellar particles (the protostellar mass function, or PMF) in our simulation and compare it with the theoretical models of \citet{mckoff10}. At present, it is not possible to measure protostellar masses, so we cannot compare with observation.
The PMF is constructed so as to produce a specified IMF when the stars reach their final masses. We base our
PMFs on the \citet{chab05} IMF, as did \citet{mckoff10}.

Neither our simulation nor the analytic models treat the formation of the first core correctly. Recently, \citet{vay16}
calculated a grid of high resolution 1D protostellar models, finding that the mass of the first core ranged
from 0.031 to 0.052 $M_{\odot}$, with an average of $0.044 M_{odot}$.
Although their models do not include a magnetic field, they find that other 3D first-core formation simulations that include magnetic fields basically match their results.
Since we do not follow the physical processes associated with the formation of the first core, our results for the PMF
are not accurate below $0.04\,M_\odot$.

Fig. \ref{fig9} shows the PMF from the simulation at two different times, when the total mass of protostars is $5 M_\odot$ (22 protostellar particles at $t=0.7\tffo$ yr,
with a maximum mass of $0.57\,M_\odot$)
and at the end of the simulation, when the total mass is $28.6 M_\odot$ (82 protostellar particles with a maximum mass of $m_u=3.07\,M_\odot$.) 
It is clear that the PMF at the first time is very different from the IMF, with a median mass of only $0.11\,M_\odot$,
compared to $0.20\,M_\odot$ for the \citet{chab05} IMF.  
By the end of the simulation, the PMF is in good agreement with the IMF--a KS test gives a $p$-value of 0.90.
This good agreement is achieved despite the fact that the maximum mass in the simulation is much less than
that in the IMF (we have adopted $m_u=120\,M_\odot$ for the IMF) because only about 2 stars are expected to have 
$m_*>3\,M_\odot$ in a sample of 82 stars. 
As shown in Fig. \ref{fig5}b, the median mass for $t>1.25\tffo$ is 
approximately constant and is close to $0.20\,M_\odot$, the IMF value.
By contrast, the mean mass is gradually increasing with time: it is $0.19\,M_\odot$ at $t=0.7\tffo$ and
$0.35\,M_\odot$ at the end of the simulation. Were we able to continue the simulation longer, and were the results
to approach the \citet{chab05} IMF, the mean mass would approach $0.55\,M_\odot$.

The accretion rate for $t>1.25\tffo$, when the median mass is approximately constant, 
is $\dot m_*\simeq 1.0\times 10^{-6}\,M_\odot$~yr\e\ as shown in 
Fig. \ref{fig6}b. As a result, the time to form a star of the median mass is about
\beq
t_{\rm sf,\,med}\simeq 2\times 10^5~~~{\rm yr},
\eeq
close to the free-fall time for the average gas above the mass-weighted median density, $\tffo=212,500$~yr (Eq. \ref{eq:tffo}).
It must be borne in mind that this estimate for the median star formation time is approximate, however, since star formation is
actively continuing in the simulation.

Five theoretical PMFs \citep{mckoff10,off11} and the Chabrier IMF are overplotted for comparison. The five PMFs are the isothermal sphere (IS) PMF, which has a constant accretion rate proportional to the temperature of the ambient gas \citep{shu77}, the competitive accretion (CA) PMF \citep{zin82,bon97}, the turbulent core (TC) PMF \citep{mck02,mck03}, and two two-component models (2CTC and 2CCA), in which the protostar is assumed to accrete as in the \citet{shu77} solution at early times and as in the turbulent core and competitive accretion models, respectively, at late times.
The five PMFs plotted in Fig. \ref{fig9}b are based on an upper limit cutoff mass of $5.6 M_\odot$.
This cutoff mass is obtained by a trial-and-error procedure that leads to slightly less than one protostar in the mass bin just above the highest non-zero mass bin.
The 2CTC and 2CCA PMFs provide better fits at the end of the simulation. Interestingly, at this time the 2CTC and 2CCA functions are similar to the Chabrier IMF. 
If we use a much larger maximum cutoff mass, such as $120 M_\odot$ for 2CCA and 2CTC models, the PMFs do not change much except for a longer tail at the high-mass end.
For masses $1\, M_\odot \leq m_*\ll m_u$, where $m_u$ is the upper limit on the stellar masses, the CA PMF has a slope identical to that of the Chabrier IMF, whereas the TC model is slightly flatter, with a slope of 1.1 \citep{mckoff10}. For the two component models, these slopes are reached only for $m_*\gg 1\, M_\odot$.
However, as discussed in Section \ref{sec:sfr} and at the end of the next section, the large intrinsic dispersion in accretion rates that we find in our simulation has not been incorporated into these models, so firm conclusions on the PMF cannot yet be drawn.

\subsection{Protostellar Luminosities}
\label{sec:plf}

\subsubsection{Analytic estimate}

For low-mass stars, the major energy source for the protostellar luminosity is the gravitational energy released by the accreting gas. The mean accretion luminosity of a protostar (neglecting the internal energy of the accreting gas) is \citep{off11}
\begin{eqnarray}
\begin{aligned}
L_{\rm acc,0} = &f_{\rm acc} \frac{G m_* \dot{m}_*}{r_*},\\
                  = &7.8 f_{\rm acc} \left( \frac{m_*}{0.25 M_{\odot}} \right) \left( \frac{2.5 R_{\odot}}{r_*} \right)\\
                     &\left( \frac{\dot{m_*}}{2.5\times 10^{-6}\, M_{\odot} {\rm yr}^{-1}} \right) \; L_{\odot},
\label{eq:lacco}
\end{aligned}
\end{eqnarray}
where $f_{\rm acc}$ is the fraction of accretion energy that is converted into radiation.
\citet{hos11} concluded that the best fit to the H-R diagram for young stellar objects is obtained for models in which
the early stages of protostellar accretion are ``hot," in that the accreting gas covers the star; 
$2.5\,R_\odot$, is a typical value of the radius for their two best-fit models, which switch from hot to cold accretion at 
$0.03\,M_\odot$ and $0.1\,M_\odot$.
We have evaluated the stellar mass in Equation (\ref{eq:lacco}) at the half-way point ($0.25\,M_\odot$) in the formation of a star of average mass, $0.5\,M_\odot$.

The accretion energy that goes into driving a wind is not converted into radiation near the protostar, so it reduces $f_{\rm acc}$ below unity. Following \citet{off09}, we adopt $f_{\rm acc} = 0.75$ as a fiducial value: Half the accretion energy is released in the disk and half when the gas transitions from the inner edge of the disk to the star, and we assume that half of the energy released in the disk is extracted mechanically (presumably in an MHD wind), so that altogether 3/4 of the accretion energy goes into radiation.
For the fiducial accretion rate of $2.5\times 10^{-6}\,M_\odot$~yr\e, the time to form a star of average mass ($0.5\,M_\odot$) is $2\times 10^5$~yr,
the same as that estimated above to form a star of median mass based on our simulation results.

For protostars that are not too massive ($m_*\la 1\,M_\odot$), the gas that enters the accretion disk is molecular. Once it
is embedded in the protostar, however, it is ionized. Let $\epsilon_I$ be the energy per unit mass required to dissociate and ionize the gas; for a gas that is 10 per cent He by number, this is $\epsilon_I=1.62\times 10^{13}$ erg g\e.
Allowing for this energy, the accretion luminosity becomes \citep{tan04}
\beqa
\begin{aligned}
L_{\rm acc} =&L_{\rm acc,0} -\dot m_* \epsilon_I,\\
                  =& 7.8\left[f_{\rm acc} \left(\frac{m_*}{0.25 M_{\odot}} \right) \left( \frac{2.5 R_{\odot}}{r_*} \right)-0.085\right]\\
                    &\left( \dis\frac{\dot{m_*}}{2.5\times 10^{-6}\, M_{\odot}\, {\rm yr}^{-1}} \right) \; L_{\odot}.
\label{eq:lacc}
\end{aligned}
\eeqa
For the fiducial accretion rate of $2.5\times 10^{-6}\,M_\odot$~yr\e, the ionization and dissociation energies subtract
$0.66\,L_\odot$ from the accretion luminosity. In this case, the average accretion luminosity is $5.2\,L_\odot$. This correction for the ionisation and dissociation energies was introduced only after the simulation was completed, so it is included in the data analysis below but was not included in the calculation of radiative feedback in the simulation.

At very early times, when $m_*/r_*<0.0085/f_{\rm acc}$ 
in solar units, $L_{\rm acc}$ can be negative, which
means that there is not enough energy available to fully dissociate and ionize the accreted gas. 
The total luminosity of the protostar is 
\beq
L={\rm max}\left(L_{\rm acc}+L_{\rm int},\, 0 \right),
\label{eq:L}
\eeq 
where $L_{\rm int}$ is the internal luminosity of the protostar, due to both nuclear energy generation and Kelvin-Helmholtz contraction. $L_{\rm int}$ is a sensitive function of the protostellar mass; for example, a 
well-sampled
Chabrier IMF truncated above a mass 
$m_u=3\,M_\odot$ gives  $L_{\rm int}=1.1\,L_\odot$, whereas $m_u=6\,M_\odot$ gives $L_{\rm int}=6.0\,L_\odot$. 
For $m_u=3\,M_\odot$, the nuclear luminosity is about 20 per cent of the fiducial value, whereas
for $m_u=6\,M_\odot$ it is 
comparable to the fiducial value.
It should be borne in mind that actual samples are finite and will not extend all the way up to $m_u$, so
the actual mean internal luminosity will be less than this.
If the energy loss due to dissociation and ionization is sufficiently large that $L_{\rm acc}+L_{\rm int}$ is negative, then
the accreted gas does not become fully ionized as it is accreted, which reduces the luminosity at later times when it does become fully ionized.

\subsubsection{Mean and median protostellar luminosities: comparison with observation}
\label{sec:tepl}

We compare the protostellar luminosities from the simulation with those of the sample of 230 young stellar objects (YSOs) in nearby molecular clouds analysed by \citet{dun13}. All the sources in this sample include data at 70 $\mu$m and at least one observation at $\lambda \geq 350\,\mu$m; the requirement of a submm/mm detection ensures that these YSOs are protostellar. The completeness limit for the entire sample is $0.2\,L_\odot$, but omission of the two sources (a tiny fraction of the total) in
IC 5146, the most distant molecular cloud in their sample, reduces the observational completeness limit to $0.04\,L_\odot$.
For the fiducial case of a protostar of mass $0.25\,M_\odot$ and radius $2.5\,R_\odot$, this luminosity corresponds to
an accretion rate of only about $2\times 10^{-8}\,M_\odot$~yr\e. 
We have applied a luminosity cutoff of
$0.04\, L_{\odot}$ to both the simulation and the \citet{dun13} data in our analysis.

In the \citet{dun13} sample, 130 sources included at least one FIR/submm photometry point at 160, 350, or 450 $\mu$m. The remaining 100 sources lacked any data between 70 and 850 $\mu$m, and sometimes between 70 $\mu$m and 1.1 mm. \citet{dun13} found that the bolometric luminosities of the 100 sources that lacked FIR/submm data are substantially lower than those of the 130 sources that had such data. From their examination of sources that have FIR/submm data, they concluded that on average the true bolometric luminosity of the sources without FIR/submm data is about 2.5 times larger than the measured value. However, there is one source for which this correction appears incorrect and which would have a significant impact on the mean luminosity: Source 57, which has $L=63\,L_\odot$ and does not have FIR/submm data, would have more than twice the luminosity of any of the stars with a completely observed spectrum if its luminosity were increased. Furthermore, the observed bolometric temperature for this source is 660 K, and there is no evidence that such hot sources emit most of their radiation
longward of 70 $\mu$m. We therefore do not apply the correction to this source. We find that the corrected mean luminosity of the observed sample is $5.5\, L_{\odot}$; the median value of the luminosity is not affected by these corrections since less than half the sources have their luminosities adjusted.
\citet{dun13} show that low luminosity sources can have observed luminosities that are significantly contaminated
(up to a few tenths of a solar luminosity) by external heating of the natal cloud, but we are unable to accurately correct for this in our simulation. \citet{dun13} also note that the observed luminosities could be affected by the orientation of the source relative to the line of sight, but this should not affect the average or median luminosities.

A very rough estimate of the internal luminosity of the \citet{dun13} sample based on our simulation results suggests that the mean internal luminosity of the protostars with $L>0.04\,L_\odot$ is about $0.6\,L_\odot$,
so that the mean accretion luminosity
of the observed sample is about $4.9\,L_\odot$. 
The internal luminosity of stars emitting the median luminosity is negligible.

\begin{figure}
\includegraphics[scale=0.35]{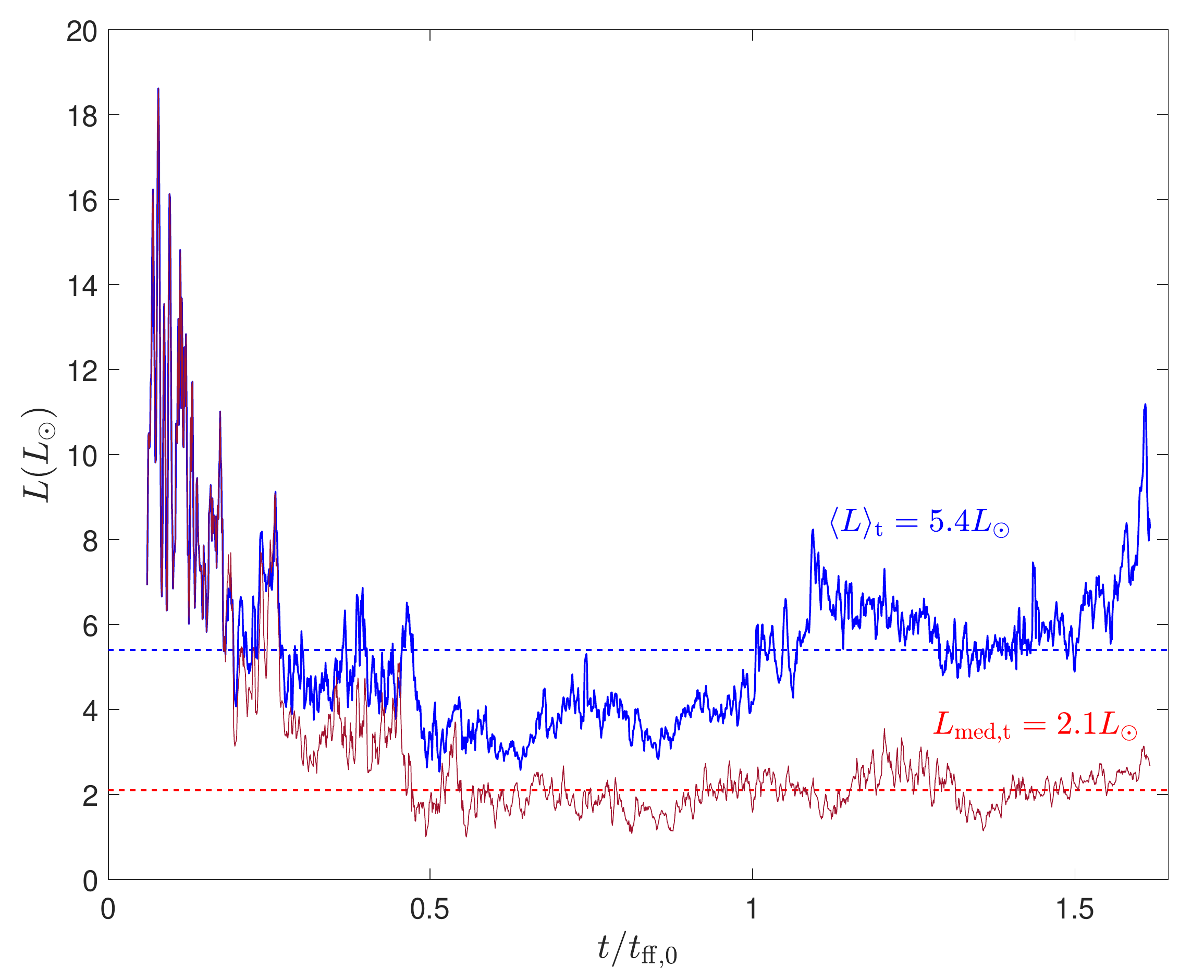}
\caption{Time evolution of the mean (thin blue) and median (thick red) luminosities of the protostellar particles. The curves are smoothed over 1000 years.
The median remains roughly constant after $0.46 \tffo$. The time averaged mean (blue dash), $\langle L \rangle_t$, and median (red dash), $L_{{\rm med},\,t}$, luminosities after $0.46 \tffo$ are 5.6 and $2.3 L_{\odot}$, and are marked by the blue and red dashed lines respectively.
\label{fig10}}
\end{figure}

In Fig. \ref{fig10}, we show the time evolution of the mean and median luminosities from the simulation, smoothed over 1,000 years. 
These luminosities are calculated from Equations (\ref{eq:lacc}) and (\ref{eq:L}) with $f_{\rm acc}=0.75$.
Our simulation includes the luminosity due to nuclear reactions in the stars, but this is only 
18 per cent of the total at the end of the simulation, when the nuclear luminosity is largest.
The initial large luminosity jump is the result of only two 
protostellar particles that exist at early times with a large accretion rate of order $10^{-5}\, M_{\odot}\, {\rm yr}^{-1}$.
The large jump near the end of the simulation is due to an increase in the accretion rate of seven protostars, six of which are in the region of the colliding filaments; three of these are in a triple star system.
Both the mean and median values quickly drop below $5 L_{\odot}$ when more particles emerge without a large initial jump in luminosity. 
After $t=0.46 \,\tffo$, the median luminosity remains about constant in time at $L_{\rm med,t} = 2.1\, L_{\odot}$.
The mean luminosity appears to be slowly increasing with time. 
After $t=0.46 \tffo$, the time-averaged mean luminosity is $\langle L\rangle_t = 5.4\, L_{\odot}$, of which $5.1\,L_\odot$ is due to accretion. These luminosities are summarized in Table \ref{tab:luminosities}.
The good agreement between the simulated and theoretical values of the average luminosity, both of which are based
on Equation (\ref{eq:lacc}), shows that the fiducial value of $\avg{m_*\dot m_*/r_*}$ agrees well with the simulation
value, which is calculated self-consistently.

One difference between the simulations and observations that we have not taken into account is FU Ori outbursts, which are accretion events that add mass to the protostar and result in substantial increases in luminosity--a factor $\sim 100$ for a period of $\sim 100$ yr \citep{har96}.
From an analysis of the observations, \citet{off11} estimated that such episodic accretion could account for about 25 per cent of the mass accreted onto a typical protostar. However, these large outbursts are so rare that there is no obvious FU Ori source in the \citet{dun13} sample. Because FU Ori outbursts are not included in the sample of observed protostars, the total amount of radiation emitted during star formation is underestimated in this sample. However, since this effect is not large and the 25 per cent estimate is uncertain, we have not made any correction to the results in Table \ref{tab:luminosities}.

The agreement of the mean and median luminosities between the simulation and observation is actually somewhat surprising, since the accretion luminosity varies as the accretion rate, which should vary inversely with the free-fall time and therefore depend on the density ($\dot m_*\propto t_{\rm ff}^{-1}\propto\rho^{1/2}$). The relatively good agreement between theory and observation therefore indicates that the densities in the observed and simulated star-forming regions are similar.
As noted above, we estimate that 
the mean accretion luminosity of the \citet{dun13} sample is about $4.9\,L_\odot$. 
This is very close to the fiducial theoretical value 
of $5.2\,L_\odot$ from Equation (\ref{eq:lacc}), and it remains close after 
correcting for the unseen FU Ori outbursts (see Table \ref{tab:luminosities}). We conclude that the combination of fiducial parameters adopted in that equation is in good agreement with observation.

\begin{table}
\caption{Protostellar luminosties with lower cutoff at $0.04\,L_\odot$}
\label{tab:luminosities}
\begin{tabular}{lcccc}
\hline
\hline
 & \multicolumn{2}{c}{Total luminosity} & \multicolumn{2}{c}{Accretion luminosity} \\
               & $\langle L \rangle$ & $L_{\rm med}$ & $\langle L \rangle$ & $L_{\rm med}$ \\
\hline
Observed    & 5.5\fnm[2]   &   1.8    & 4.9\fnm[3]     &      1.8            \\
Simulation \fnm[1]  &      5.4    &   2.1    &        5.1     &      2.1    \\
Theory \fnm[4]     &       --     &    --     &         5.2     &      --      \\
\hline
\hline
\end{tabular}

\begin{flushleft}
\fnt{1} {$^{\rm 1}$ Time-averaged for $t>0.46\tffo$.} \\
\fnt{2} {$^{\rm 2}$ This value is obtained from applying a correction factor of 2.5 to all but one of the 100 sources in \citet{dun13} that are lacking FIR/submm data.  
See Section \ref{sec:tepl} for discussion.} \\
\fnt{3} {$^{\rm 3}$ Based on an estimated $L_{\rm int} \sim 0.6 L_{\odot}$.} \\
\fnt{4} {$^{\rm 4}$ From equation (\ref{eq:lacc}) using fiducial values of $m_*$, $\dot m_*$, $r_*$ and $f_{\rm acc}$.} \\
\end{flushleft}
\end{table}

\begin{figure}
\includegraphics[scale=0.35]{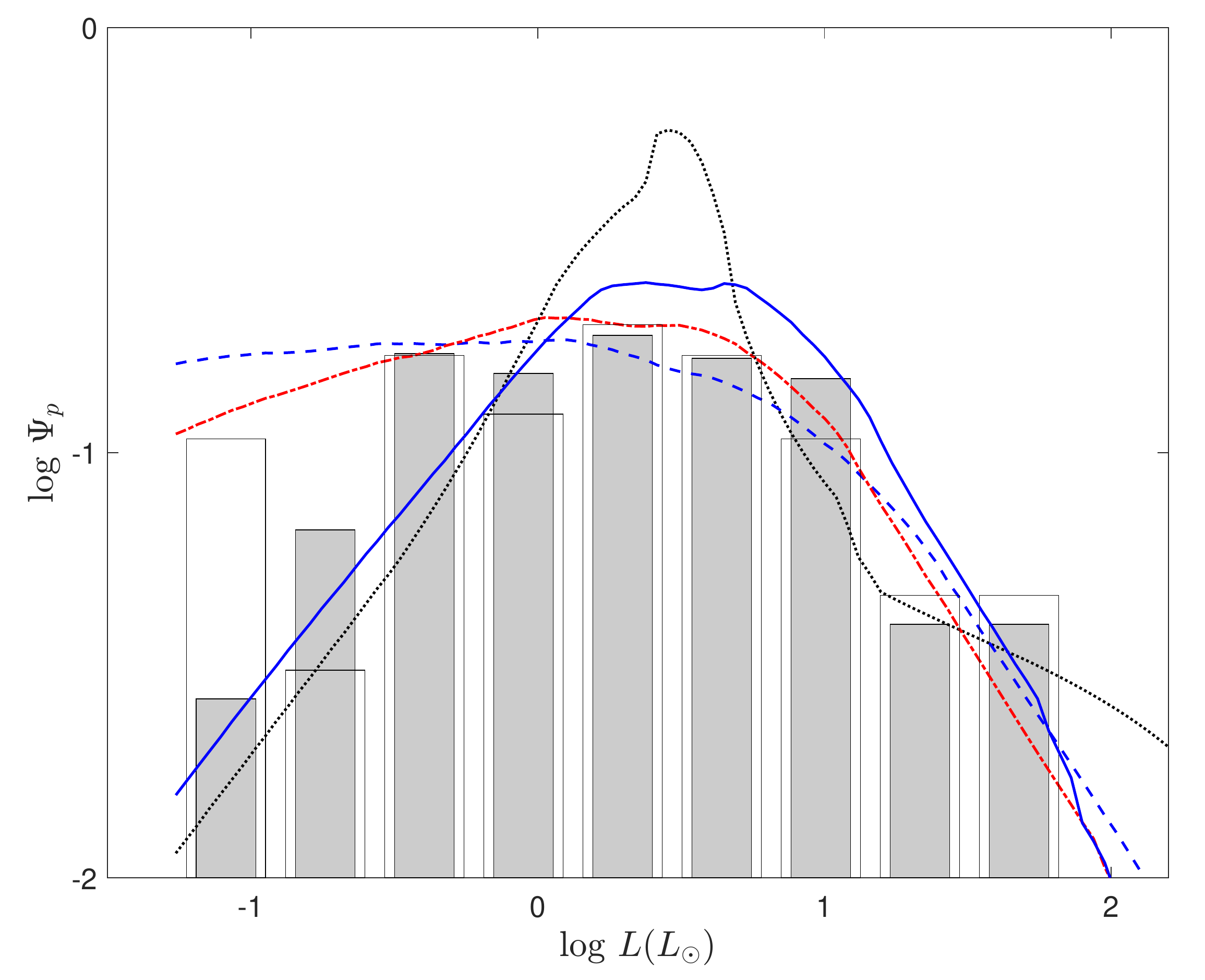}
\caption{Comparison of PLFs from the simulation 
(empty histogram) with the FIR/submm corrected sources from \citet{dun13} (shaded histogram) and with four theoretical PLF models: IS (black dotted), CA (blue dash), TC (red dot-dash), and 2CTC (blue solid)
\citep{off11} for the scenario of untapered, non-accelerating accretion.
The data from the simulation is at $t\simeq 1.3\tffo$, the last time at which the total luminosity is about the same as the average value from \citet{dun13}, $6.0\,L_\odot$.
The histograms and curves are normalized to have the same area.
The model that gives the best fit to the observations is the 2CTC model.
\label{fig11}}
\end{figure}

\subsubsection{The protostellar luminosity function}

In Fig. \ref{fig11}, we plot the observed PLF from \citet{dun13}, omitting sources with $L<0.04\,L_\odot$ and increasing the luminosity of the sources without FIR/submm data by a factor 2.5, as noted above; a snapshot of the simulated PLF; 
and the PLFs for four of the five models considered above for the PMF. 
Fig. \ref{fig10} shows that the mean luminosity of the simulated PLF is gradually increasing in time after an initial decline. To make the best comparison with observation, we use the simulated PLF
at the last time $(\simeq 1.55 \tffo)$
when the mean luminosity is the same as that from \citet{dun13}.
The total mass of protostars in the simulation at that time is $26\,M_\odot$.
The theoretical PLFs are calculated with $T=10$~K, $\Sigma_{\rm cl}=0.53$ g cm\ee, and a competitive accretion rate that is 3.2 times the isothermal accretion rate (see \citealp{off11}).
There are 65 protostars with $L>0.04\,L_\odot$ at the time of the snapshot in Fig. \ref{fig11}.
The mean and median luminosities of the simulated sources at the time of this snapshot are $5.5\,L_\odot$ and $1.9\,L_\odot$, very close to the time-averaged values in Fig. \ref{fig10}.
The standard deviation of the protostellar luminosities is 0.76 dex at this time, comparable to the 0.65 dex in Dunham et al.'s sample. This dispersion in the luminosity includes the effects of both the fluctuations in the accretion rate of individual protostars and the dispersion in the mean accretion rates of the collection of protostars (Section \ref{sec:sfr}), as well as the dispersions in protostellar masses and radii.

Fig. \ref{fig11} also includes four theoretical model curves from \citet{off11}.
The model PLFs are computed based on an upper mass of $5.6 M_{\odot}$ at $1.3 \tffo$
Compared to the plots in \citet{off11}, the model PLFs in Fig. \ref{fig11} show more features because the time evolution of the protostellar radii in \citet{off11} is smoothed before integrating to obtain the luminosity function, whereas in this work we do not apply any smoothing to the radii.
We are not able to consider models in which the accretion rate tapers off as the star reaches its final mass because we had to stop the simulation before most of the stars stopped accreting. The model that fits best is the two-component turbulent core model (2CTC).
We did not have the data required to compute the PLF for
the two-component competitive accretion model (2CCA), but it should be similar to that of the 2CTC model.
However, as noted in Section \ref{sec:pmf} above and discussed further below, 
the broad dispersion in accretion rates that we find are not
included in the analytic models, so firm conclusions on the accretion model cannot be drawn.

The PLF has recently been considered in simulations by \citet{pad14}. Their simulations did not include radiative feedback, which has been shown to be important in setting the characteristic mass of the IMF \citep{bat09,off09,kru16}, nor outflows, which also affect the IMF (e.g., \citealp{han12}); our simulation includes both effects. On the other hand, they included the effects of heating by the interstellar radiation field, which we have not. The parameters of their simulation are quite similar to ours: The thermal Mach number is 10, the mass is about $3000\,M_\odot$, and the mean density is about 800 cm\eee\ (compared to our values for the same quantities of 10, 3110 and 960). This is not surprising since one is naturally led to these values, as discussed in Section \ref{sec:sim}. Their initial magnetic field was significantly weaker than ours, with $\ma=5$, compared to our value of unity. However, turbulence amplifies this field so that during the star-forming part of the simulation it is not that much less than in our simulation. Their previous work \citep{pad12} suggests that the star formation efficiency decreases as the magnetic field strength increases until $\ma=5$ and then 
increases as the field increases further so that $\ma$ drops below 5. 
This suggests that the simulation in \citet{pad14} has a smaller mass accretion rate than our zoom-in simulation, which has $\ma=1$, although we cannot confirm that directly. Another difference between their simulation and ours is that they set
$f_{\rm acc}=0.5$ instead of 0.75, as we did; however, after allowing for episodic accretion, we have $f_{\rm acc,\,eff}=0.56$, which is quite close to their value of $f_{\rm acc}$. They study their data at 2.6 Myr, and at that time $\eff$ is somewhat more than 0.06, similar to the value of $\eff$ at the end of our simulation. They reported that the mass accretion rates onto their sink particles are in the range of $10^{-5}-10^{-6} M_{\odot}$ yr$^{-1}$, 
also similar to that for most of our protostellar particles. From the PLF in their figure 11 we estimate that the mean and median luminosities in their simulation at 2.6 Myr are about 10.7 and $0.75\, L_{\odot}$. The mean value is very similar to ours
at the end of the simulation (although larger than the time-averaged value),
but the median is smaller than both our value and the observed value \citep{dun13}. They have more low luminosity particles than we do, which we attribute to their omission of radiative feedback, which suppresses fragmentation \citep{bat09,off09,kru16}.
They note that their PLF fits the observations by \citet{kry12} quite well.
\citet{dun13} note that a number of the low luminosity sources in \citet{kry12} do not have detected (sub)millimeter emission, so they either are associated with low-mass cores ($\la 0.5-0.8\,M_\odot$) or are not protostars; clearly, more sensitive observations are needed to resolve this discrepancy.

\begin{table*}
\caption{Stellar multiplicity at the end of the simulation}
\label{tab:multiplicity}
\begin{tabular}{llllllllll}
\vspace{-0.3cm}\\
\hline
\hline
\vspace{-0.2cm}\\
Separation Range & Projection\fnm[1] & S & B & T & Q & $>4$ & MF & CF & Observed CF (MF)\\
\hline
$100-1000$ au & 3D & 41 & 4(2) & 2(0) & 0 & 0 & 0.11 & 0.17 \\
              & x  & 27 & 7(1) & 3(0) & 1(1) & 1(0) & 0.29 & 0.42 \\
              & y  & 38 & 5(2) & 1(0) & 1(0) & 0 & 0.16 & 0.22 \\
              & z  & 41 & 2(1) & 3(0) & 0 & 0 & 0.11 & 0.17 \\
              & 2D & & & & & & $0.19\pm0.09$ & $0.27\pm0.13$ & 0.17 (0.17)\fnm[3] \\
              \\
$110-1400$ au & 3D & 36 & 6(2) & 1(0) & 1(1) & 0 & 0.18 & 0.25 \\
              & x  & 25 & 8(1) & 1(0) & 1(1) & 1(0) & 0.31 & 0.50 \\
              & y  & 31 & 8(2) & 1(0) & 0 & 1(1) & 0.24 & 0.35 \\
              & z  & 36 & 4(1) & 2(0) & 1(1) & 0 & 0.16 & 0.26 \\
              & 2D & & & & & & $0.24\pm0.08$ & $0.37\pm0.12$ & $0.27\pm0.06$\fnm[4] \\
              \\
$300-2000$ au & 3D & 34 & 7(0) & 0 & 1(1) & 0 & 0.19 & 0.24 \\
              & x  & 26 & 6(0) & 1(0) & 2(0) & 0 & 0.26 & 0.40 \\
              & y  & 31 & 8(0) & 0 & 1(1) & 0 & 0.23 & 0.28 \\
              & z  & 33 & 6(0) & 2(0) & 0 & 0 & 0.20 & 0.24 \\
              & 2D &  & & & & & $0.23\pm0.03$ & $0.31\pm0.08$ & $0.18\pm0.04$\fnm[5] \\
              \\
$100-5000$ au & 3D & 24 & 8(2) & 3(0) & 0 & 1(0) & 0.33 & 0.53 \\
              & x  & 23 & 7(1) & 2(0) & 0 & 2(0) & 0.32 & 0.59 \\
              & y  & 20 & 10(2) & 3(0) & 0 & 1(1) & 0.41 & 0.62 \\
              & z  & 24 & 7(1) & 3(0) & 0 & 1(0) & 0.31 & 0.54 \\
              & 2D & & & & & & $0.35\pm0.06$ & $0.58\pm0.04$ & 0.46 (0.35)\fnm[6] \\
\hline
\hline
\end{tabular}
\begin{flushleft}
\fnt{1} {$^1$ The 3D separation range or a 2D-projection separation range along the three cardinal axes}.\\
\fnt{2} {$^2$ The bracketed values are the numbers of multiples that are gravitationally bound.}\\
\fnt{3} {$^3$ Class I phase CF (MF) from \citet{tob16}}.\\
\fnt{4} {$^4$ Class I phase CF in the Taurus and Ophiuchus clouds from \citet{duc04}.\\
\fnt{5} {$^5$ Class I phase CF in the Taurus, Ophiuchus, Serpens, Perseus, Chamaeleon I and II clouds from \citet{hai04}.}\\
\fnt{6} {$^6$ Class I phase CF (MF) from \citet{con08}}.}\\
\end{flushleft}
\end{table*}

\citet{pad14} made the important point that the broad dispersion in the mass accretion histories that they found has a profound effect on the PLF: Whereas models with fixed, time-independent accretion rates, such as the isothermal model, are clearly ruled out by the data, models with accretion rates that fluctuate about a mean, and that have a large dispersion in the mean accretion rates, are consistent with the data. As discussed in Section \ref{sec:sfr}, we too find that the accretion rates fluctuate about a mean until in some cases they begin declining, and furthermore that there is a large dispersion in the mean accretion rates. As a result, they conclude that star formation histories are better described by constant-rate models, contrary to \citet{off11}, who concluded that the histories are better described by models in which the star formation time depends only weakly on the stellar mass (constant-time models). 
Models for massive star formation \citep[e.g.][]{mck03,mur15} typically have
an accretion rate that increases with time and fall into the constant-time class of models. We did not form any  stars above $m_*\sim 3\,M_\odot$ in our simulation and \citet{pad14} did not form any above $10\,M_\odot$; neither found evidence for protostellar accretion rates that systematically increase with time. By contrast, \citet{lee15} did form more massive stars in their simulation, and found that the accretion rate for the most massive stars in their simulation increased approximately linearly in time. Note that the two-component models
(2CTC and 2CCA) that fit our simulations best have constant accretion rates at low mass and time increasing rates at high mass, so they have both a constant-rate component and a constant-time component. Further work is needed to extract the information contained in observed protostellar luminosity functions.

\citet{ken90} pointed out that the observed luminosities of Class I sources in Taurus-Auriga are considerably less than the value implied by Equation (\ref{eq:lacc}): the mean and median luminosities in their sample are 2.6 and 1.6 $L_\odot$, respectively. On the other hand, they inferred an accretion time of $1-2\times 10^5$ yr, corresponding to an accretion rate for a star of mean mass ($0.5 M_\odot$) of $2.5-5 \times 10^{-6} \,M_{\odot}\, {\rm yr}^{-1}$. As a result, Equation (\ref{eq:lacco}) with $f_{\rm acc}=1$ and $r_*=2.5\,R_\odot$ implies an average luminosity of 
$7.8-15.6\, L_\odot$, 3-6 times the observed value they found.
Note that their estimate of the accretion rate was based on an estimate of the protostellar lifetime, and is thus
the average accretion rate, $m/\tsf$, which corresponds approximately to the average luminosity. 
(They compared this theoretical average value with the observed mode of the luminosity distribution, which made the discrepancy between theory and observation appear even larger.) The results of \citet{dun13}, which are based on a larger sample of protostars with greater wavelength coverage, give a comparable value for the median luminosity ($1.8\,L_\odot$) but a mean luminosity that is more than twice as large ($5.8\,L_{\odot}$). 
Allowing for the energy needed to drive the observed protostellar outflows (estimated as $1-f_{\rm acc}=0.25$), the
predicted average accretion luminosity (Eq. \ref{eq:lacc}) is about 1-2 times the observed value, 
and FU Ori outbursts reduce this
to about 0.6-1.3 if they account for 25 per cent of the accreted mass, as estimated by \citet{off11}. Given that
the average protostellar luminosity depends on
the accretion rate, $\dot m_*$,
the protostellar radius, $r_*$, the radiative efficiency, $f_{\rm acc,\, eff}$, and the accretion histories of the protostars,
all of which are uncertain, we conclude that there is no evidence that protostellar luminosities are less than expected. 
Recall that our numerical results, which self-consistently evalulated all these quantities except $f_{\rm acc,\,eff}$, are also consistent with the observations of \citet{dun13}. 
Our conclusion that there is no protostellar luminosity problem agrees with that of
\citet{off11} and \citet{pad14}.

\subsection{Multiplicity}
\label{sec:multiplicity}

\begin{table}
\caption{Gravitationally bound systems as function of 2D projection spacing}
\label{tab:gbound}
\begin{tabular}{lllll}
\vspace{-0.3cm}\\
\hline
\hline
\vspace{-0.2cm}\\
Separation Range (au) & $N_x$ & $N_y$ & $N_z$ & $\langle \% {\rm bound} \rangle$\\
\hline
Binary \\
\hline
100 - 200   & 1(1)\fnm[1] & 1(1) & 1(1) & 100 \\
200 - 400   & 1(0) & 1(1) & 0    &  50 \\
400 - 800   & 1(0) & 1(0) & 1(0) &   0 \\
800 - 1600  & 4(0) & 2(0) & 1(0) &   0 \\
1600 - 3200 & 0    & 2(0) & 4(0) &   0 \\
3200 - 6400 & 0    & 2(0) & 0    &   0 \\
\hline
Triple \\
\hline
100 - 200   & 0    & 0    & 0    &   0 \\
200 - 400   & 0    & 0    & 0    &   0 \\
400 - 800   & 0    & 0    & 1(1)\fnm[2] &   0 \\
800 - 1600  & 1(0) & 0    & 1(0) &   0 \\
1600 - 3200 & 1(0) & 1(0) & 1(0) &   0 \\
3200 - 6400 & 0    & 0    & 0    &   0 \\
\hline
Quadruple \\
\hline
100 - 200   & 0 & 0    & 0 &   0  \\
200 - 400   & 0 & 0    & 0 &   0  \\
400 - 800   & 0 & 0    & 0 &   0  \\
800 - 1600  & 0 & 0    & 0 &   0  \\
1600 - 3200 & 0 & 0    & 0 &   0  \\
3200 - 6400 & 0 & 1(0) & 0 &   0  \\
\hline
$>4$ \\
\hline
100 - 200   & 0 & 0    & 0    &   0  \\
200 - 400   & 0 & 0    & 0    &   0  \\
400 - 800   & 0 & 0    & 0    &   0  \\
800 - 1600  & 1 & 0    & 0    &   0  \\
1600 - 3200 & 1 & 0    & 0    &   0  \\
3200 - 6400 & 0 & 2(0) & 1(0) &   0  \\
\hline
\end{tabular}
\begin{flushleft}
\fnt{1} {$^1$ The bracketed values are the numbers of multiples that are gravitationally bound.}\\
\fnt{2} {$^2$ This triple has a nearby protostar just below the luminosity cutoff, so it is actually part of a bound quadruple.}\\
\end{flushleft}
\end{table}

\subsubsection{Method}

Here we study the multiplicity of the stellar systems in our simulations. 
We define the multiplicity fraction, MF \citep{hub05}, and the companion fraction, CF (e.g., \citealp{hai04}), as
\beq
MF \equiv \frac{B+T+Q+...}{S+B+T+Q+...},
\label{eq:mf}
\eeq

\beq
CF \equiv \frac{B+2T+3Q+...}{S+B+T+Q+...},
\label{eq:cf}
\eeq
where $S, B, T$, and $Q$ are the numbers of single, binary, triple, and quadruple systems. 

We determine which stellar systems are multiple both in 2D, using the projected spatial separations of the stars as observers do, and in 3D. Since the size of the finest cell in the simulation is about 28 au, we can resolve binaries down to about 4 times that, or 100 au. Any group of protostars separated by less than 100 au (3D) or less than the projected distance of 100 au (2D) is grouped into a single star. 
After applying the 100 au requirement in 3D, there are 64 ``stars" left. Of the 12 stellar systems containing more than one star within 100 au, 8, 2, and 2 have two, three or four stars within that radius. The dynamics of these close binaries is inaccurate, since gravity is implemented with a softening radius of one cell.
We set a luminosity cutoff of $0.011 L_{\odot}$ since that is the minimum in the four surveys we compare with. (However, we do include the gravitational effects of fainter stars below.) Our procedure for constructing the multiples in the simulation is analogous to that of \citet{bat09} and \citet{kru12b}: 
We determine the 2D and 3D separations of all the stars and form a binary of the two stars with the closest (projected) separation provided that it is within the specified range. 
Generalizing the procedure of \citet{tob16}, 
the protostar with the larger luminosity is designated as the primary, and the binary is located at the position of the primary. This procedure is then repeated. 
Higher order multiple systems are formed if one or both members of the ``binary" formed in this procedure are multiple.
The largest system we found within our maximum range of 5000 au consists of 7 protostars. 
When checking if two multiples can form a new group within the specified range of separations, we determine the primary of the proposed system and require that all members of this system be within the specified range of the primary; if not, the two multiples are not joined.

\subsubsection{Results}

Table \ref{tab:multiplicity} shows the results for the stellar multiplicity at the end of the simulation with separation ranges from several sets of observations of Class I protostars. The protostars from the simulation are probably a mixture of Class 0 and I, but \citet{tob16} have shown that the MF and CF values for Class 0 protostars are usually close to those of Class I protostars.
We compare with observed samples that have a minimum separation of at least 100 au
(generally set by the angular resolution of the observations), which is
the same as that in our simulation (set by the resolution of our calculations).
Most of these papers gave only the CF, so we focus on that.
The CFs at the end of the simulation match reasonably well (within a factor 2) with the observations of different ranges of separation.
The CF values from the simulation are larger than the observed values; this could be due to multiples with more than 4 members,
which appear to be more common in the simulation than in the observations.

In the simulations, 
both the MF and CF generally increase as the separation range increases except for the range 300 - 2000 au.
We can see a clear projection effect on both the CF and MF values along different lines of sight: Since the dark cloud filament lies close to the $x$-axis of the simulation box, projections along the $x$-axis lump more protostars together than along the other two axes.  As a result, both the MF and CF values along the $x$-axis are up to a factor 2 larger than those from the other two projections.

We also determine the number of multiples identified using the projection method that are truly gravitationally bound and list the numbers in parentheses. 
To determine if a multiple is bound, we first require that the kinetic energy of the system, as measured in the center of mass frame, be less that the gravitational potential energy. We then require that the acceleration of each member of the multiple due to other members of the multiple exceed the tidal acceleration (measured with respect to the primary) due to nearby protostellar particles and due to the gas. 
Most identified multiples are found to be unbound because they violate the tidal criterion, mainly due to other protostellar particles. 

We further examine the gravitationally bound systems as a function of 2D projected distance in Table \ref{tab:gbound}, which gives the numbers of identified multiples and the percentage that are gravitationally bound. We can see the fraction of gravitationally bound systems identified from projected separations is extremely low, except for systems within 400 au;
4 out of 5 binaries 
with projected separations less than that are bound.
There is an exceptional case in which a triple listed in Table \ref{tab:gbound} is actually part of a bound quadruple, but the luminosity of the fourth star is just below the luminosity cutoff we adopted in forming multiples. This system has a projected separation in the $y-z$ plane in the range between 800-1600 au;
it is not seen in other projections because the members of the quad are grouped into other multiples in those projections.
With up to 6400 au projected separation, we find only one septuple, which is unbound.  
(As noted above, the maximum separation from the primary within this system is actually less than 5000 au.) 
Two members of this system have separations much greater than 6400 au in 3D.

\begin{figure}
\includegraphics[scale=0.4]{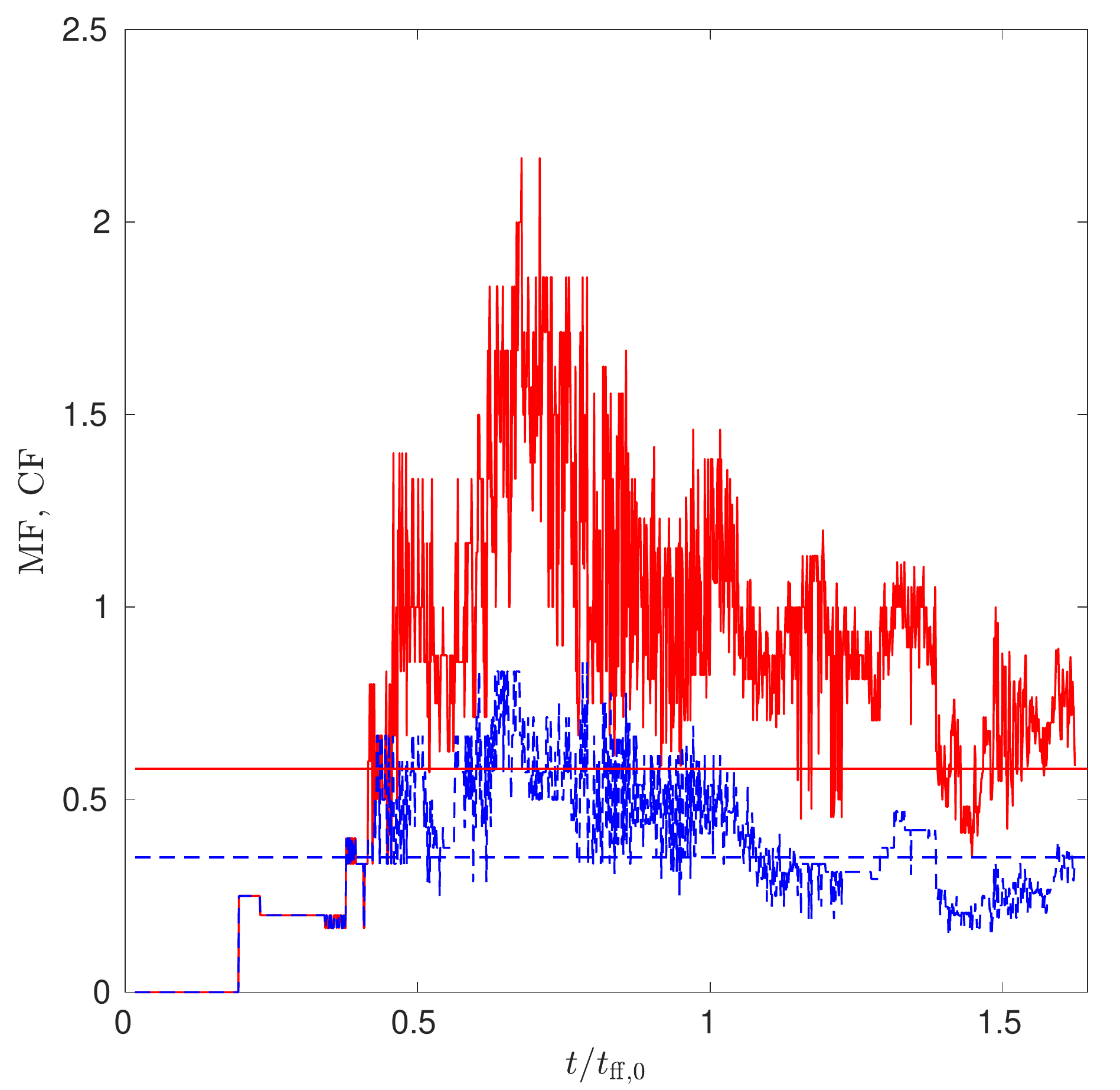}
\caption{Time evolution of multiplicity fraction $MF$ (blue dashed) and companion fraction $CF$ (red solid) in the range $100-5000$ au. The fractions reach a peak around $0.65\tffo$ and slowly decline in time. The horizontal lines, with corresponding line-type and color, are the MF and CF values at the end of the simulation. See Table \ref{tab:multiplicity} for quantitative comparison with observations.
\label{fig12}}
\end{figure}

Fig. \ref{fig12} shows the change of MF and CF with time from the simulation for the case of $100-5000$ au. Although the first protostellar particle forms soon after the beginning of the zoom-in simulation, no multiple system forms until after $0.3 \tffo$. Both MF and CF reach a peak at about $0.75 \tffo$ and slowly decline thereafter. The horizontal lines in the figures show the values of the MF and CF, 0.35 and 0.58, for the range 100-5000 au at the end of the simulation.
(see Table \ref{tab:multiplicity}). 
The decline of CF with time in the simulation is consistent with the observed decline of CF for Class I sources with stellar age as measured by the spectral index \citep{con08}. 
For example, the solar-type main-sequence CF in the range of $100-4500$ au is 0.16 \citep{duq91}, significantly smaller than the observed Class I CF of 0.46.

\subsection{Protostellar Outflows}
\label{sec:outflow}

\begin{figure*}
\includegraphics[scale=0.51]{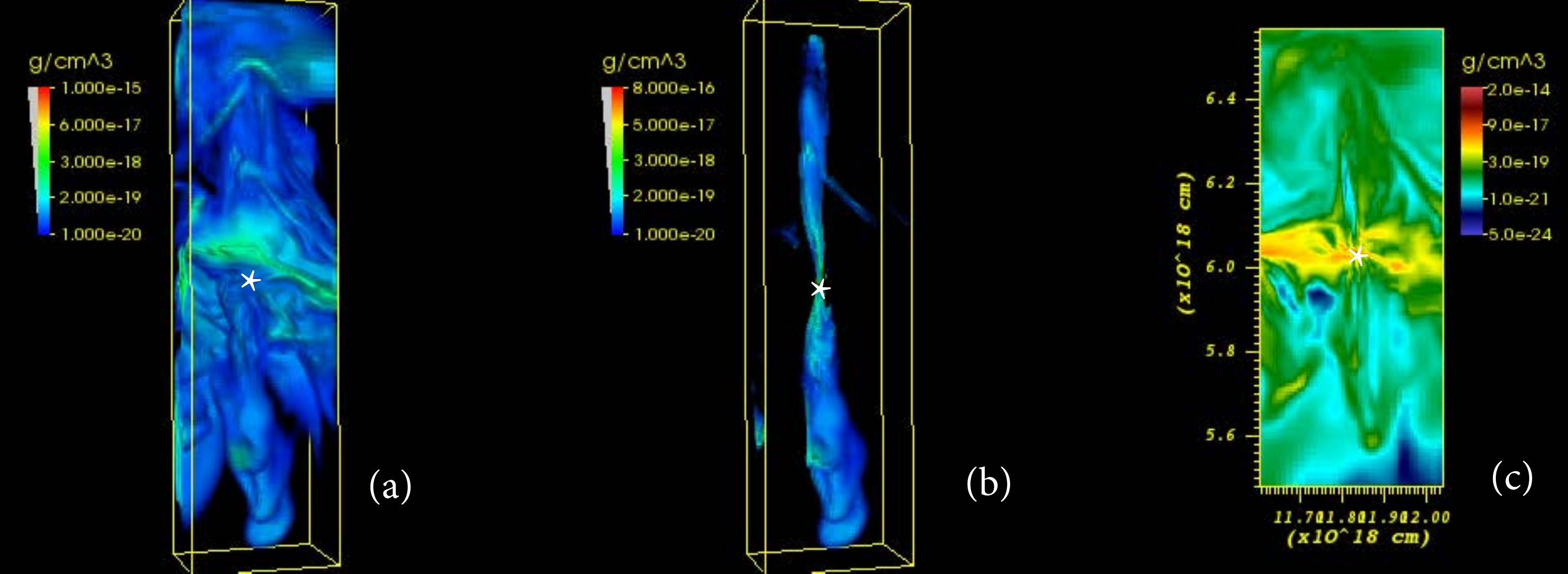}
\caption{Outflows from an unresolved ($< 100$ au)
protostellar binary
observed along the $z$-axis; the $y$-axis is vertical: (a) volume rendering of all the gas around a collimated outflow from the protostellar binary (marked by the white star symbol), (b) volume rendering of the outflow gas, defined as having a speed $> 5$ km s$^{-1}$, 
showing the collimated outflow, (c) density slice in the $x-y$ plane through the binary showing the all the gas around the outflow. The height of the bounding boxes is about 0.36 pc in both the (a) and (b) panels and is along the $y$-axis. The dense gas near the middle 
of panels (a) and (c)
is the cloud filament.
\label{fig13}}
\end{figure*}

\begin{figure*}
\includegraphics[scale=0.35]{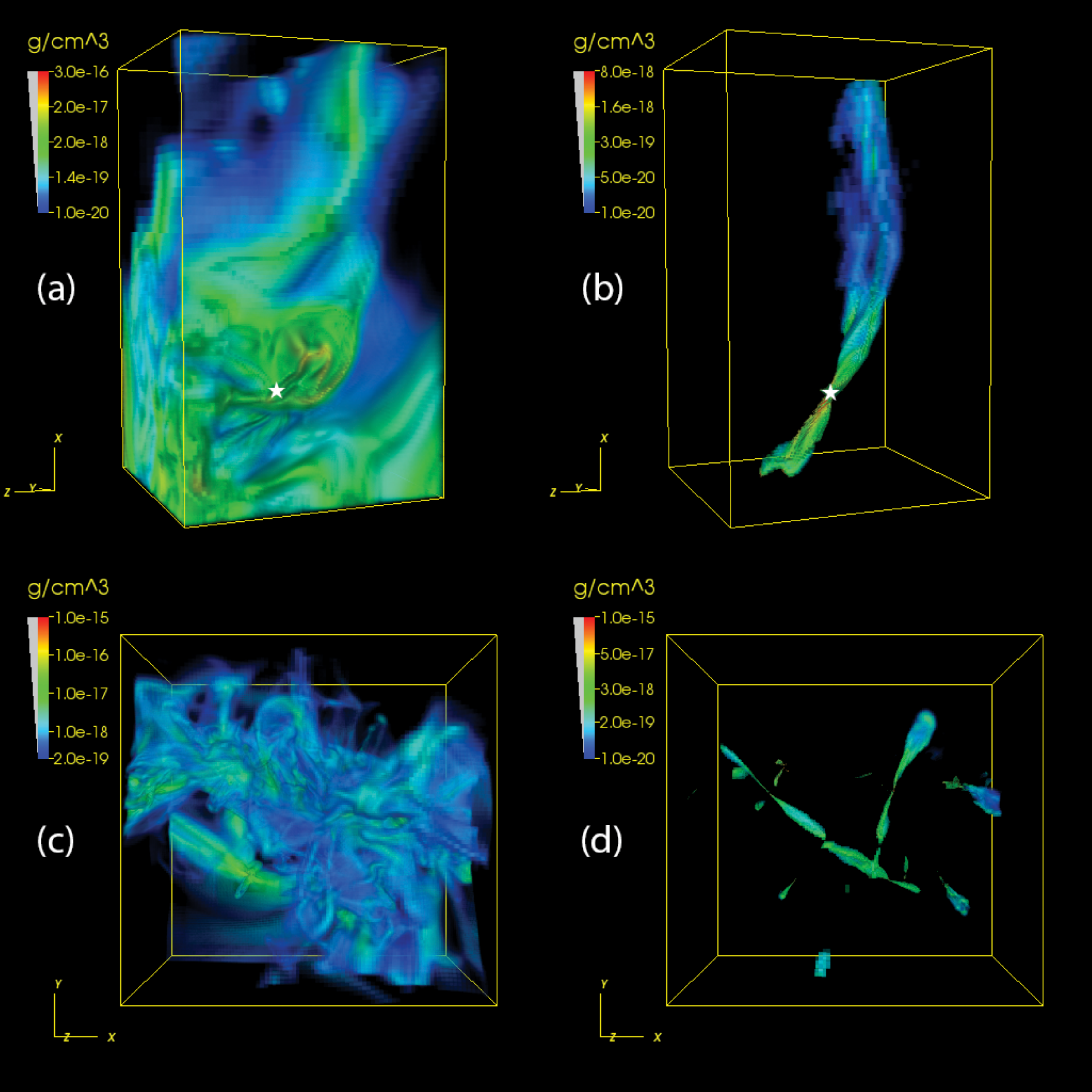}
\caption{(a) Volume rendering of all the gas around the outflow from an unresolved ($< 100$ au) binary (marked by the star symbol), (b) volume rendering of outflow gas with speed $> 5$ km s$^{-1}$ showing the short downward outflow and the longer and curved upward outflow, (c) volume rendering of all the gas in a $\sim0.2$ pc cubical region surrounding a group of protostars, (d) volume rendering of the outflow gas with speed $> 5$ km s$^{-1}$ from a group of protostars. The height of the bounding box in panels (a) and (b) is $\sim0.36$ pc.
\label{fig14}}
\end{figure*}

Highly collimated outflows \citep[e.g.][]{lad85,bac96,leecf15} inject a significant amount of kinetic energy into the surrounding environment, driving turbulence and influencing both the SFR and the final masses of the protostars \citep[e.g.][]{nak07,han12,lih15}. The physical properties of protostellar outflows are discussed in the recent review by \citet{bal16}.
Although there is no clear observational evidence that delineates the exact launching mechanism of the protostellar outflows, they are believed to be of magnetic origin, such as a disk-wind and/or an X-wind \citep{kon00,shu00}.

\subsubsection{Method}

The highest resolution in our zoom-in simulation (28.6 au) is not adequate to resolve either a disk-wind or an X-wind, each of which gets most of its energy on sub-au scales. We therefore use the subgrid model of protostellar outflows described in \citet{cun11}, which is based on the analytic model of \citet{mat99}, to reproduce the effect of protostellar outflows on the surrounding gas. Following \citet{cun11}, the fraction of the accreted mass ejected in the outflow is 0.21, and the characteristic opening angle of the flow is 0.1 radian. The outflow is injected from a spherical shell extending from the edge of the accretion zone, which is four finest cells (114 au)
from the protostellar particle, to 8 finest cells (230 au). 
Therefore, the kinetic energy injected into the ISM depends on the mass accretion rate onto the protostar. We set the protostellar outflow to begin when the mass of the protostar particle reaches 0.05 $M_\odot$.
The direction of the outflow is along the direction of the angular momentum of the protostar, which is 
proportional to the total angular momentum of the gas accreted
onto the protostar \citep{dru15}.
A more realistic model for the outflow would be to launch it normal to the protostellar disk, but none of the protostars in our simulation have disks, due to the low resolution.
We cap the maximum velocity at 100 km s$^{-1}$ in order to prevent an extremely small time step.

\subsubsection{Results: characteristics of outflows}

Outflows are visible in most of the protostars in the simulation, except for a few that are too young or that have accretion rates that are too small. Because most of the protostellar particles (including the ones separated by less than 100 au) are in multiple systems, their outflows are usually mingled together.  For well separated singles and binaries, long collimated outflows are clearly visible.  If the outflows of the two stars in a binary are in almost the same direction, the outflows appear like those from single stars. Considering the seven ``binaries" (none of which are bound) 
separated by more than 400 au, so that their accretion zones are separated by about the width of one accretion zone, and less than 2000 au, we find that their relative orientations are approximately random, and well-separated outflows are often visible.
The spatial extent of an outflow varies, depending on the environment and the accretion rate of the protostellar source.

In the following, we describe two examples of well-defined outflows, each of which comes from an unresolved ($< 100$ au) protostellar binary. Fig. \ref{fig13}, shows an outflow from a binary separated by 37 au that happens to be aligned almost along one of the cardinal axes and is well separated from other outflows, which simplifies the visualization. We define the outflow gas as gas with a velocity larger than 5 km s$^{-1}$. 
The outflow is strong enough to punch through the local dense gas.
The upward and downward outflows in panel (b) are roughly symmetrical, with
aspect ratios $>7$. 
In Fig. \ref{fig13}b,c we can see that the outflows have more than one dense blob along the jet-like outflows. Some of the structure in the outflows is due to the fact that the outflows from the two protostars are slightly offset from each other.
A recent study by \citet{off16} found that the outflows from the members of a binary can be misaligned as the result of turbulent fragmentation.

In Fig. \ref{fig14}a and b, we show an example of asymmetrical collimated outflows from a binary separated by 56 au. The downward outflow is much shorter than the upward one because the gas is denser for $r\ga 0.1$ pc below the star than above it.
The upward outflow curves because the dense filament apparent in the upper part of panel (a) deflects the outflow in panel (b) so that it propagates just to the left of the filament.  
Curved outflows are observed in clouds such as the NGC1333 region \citep{bal07}.
Therefore, the appearance and extent of outflows depend not only on the strength of  the outflows but also on the ambient gas distribution around the protostar. 

\begin{figure}
\includegraphics[scale=0.35]{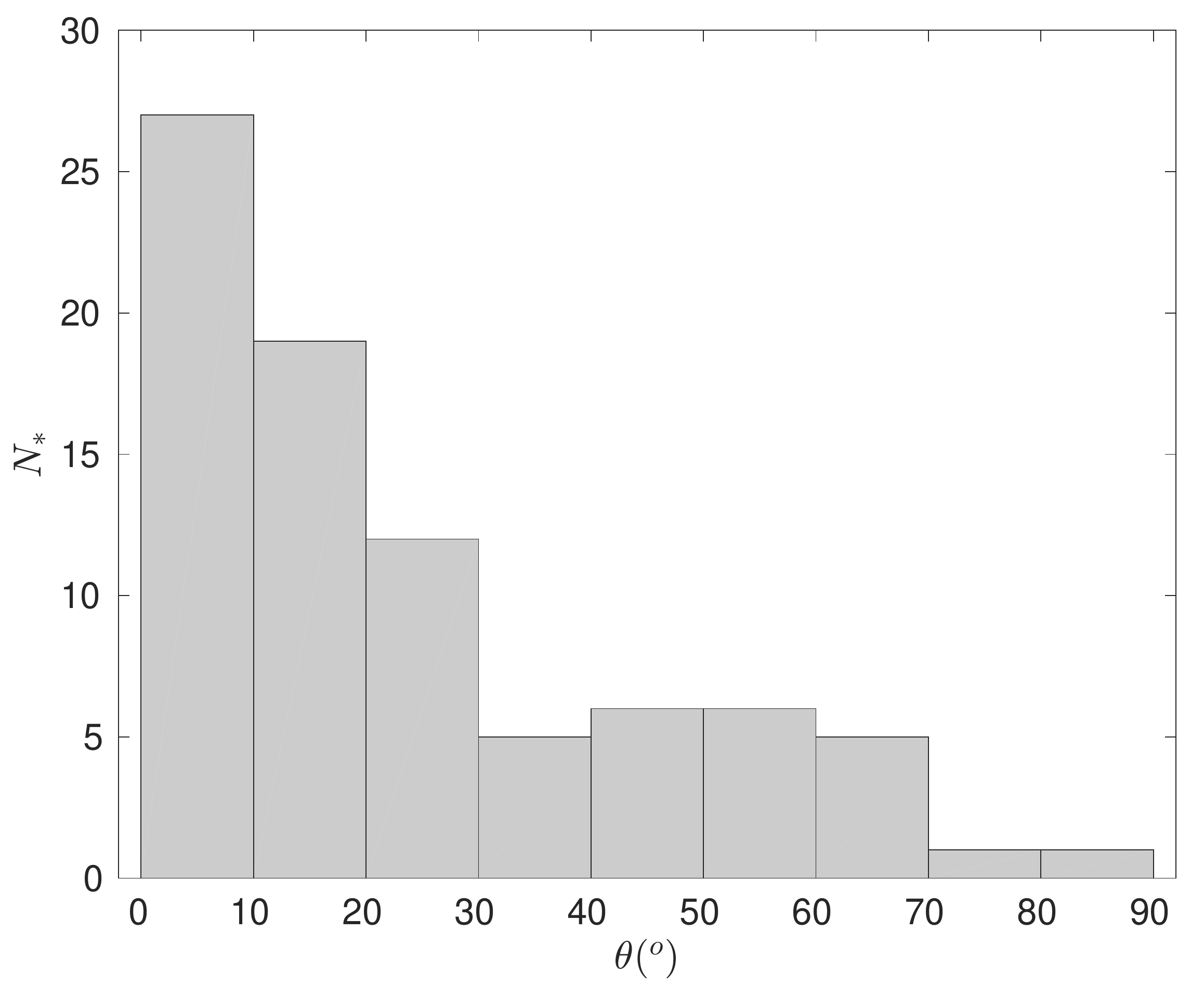}
\caption{Distribution of angles between outflows launching direction and the large scale mean magnetic field orientation. 83 per cent of the outflows are within $45^{\circ}$ from the large scale mean field direction.
\label{fig15}}
\end{figure}

In Fig. \ref{fig14}c and d, we show a cubical region $\sim0.2$ pc in length with a number of embedded protostars. The outflows from these protostars point in different directions. 
The protostars in this region are relatively young compared to the first two examples. In time, these outflows could interact with each other because of their close proximity.
Figure \ref{fig15} shows the outflow launching direction with respect to the large scale mean magnetic field direction, which is the same as the direction of the initial uniform field.
Note that the mode of this distribution is at very small angles ($<10^\circ$), corresponding to a direction approximately normal to the long filament.
Most of the outflows are relatively close to the mean field direction--only 17 per cent are more than $45^{\circ}$ from the mean-field direction.
It should be noted, however, that the outflows in our simulation were launched along the direction of the stellar angular momentum; had we launched them normal to the disks, there most likely would have been a greater dispersion in orientation based on the results of \citet{dru15}. As noted above, the resolution of our simulation is too low to implement such a model.
\begin{figure}
\includegraphics[scale=0.35]{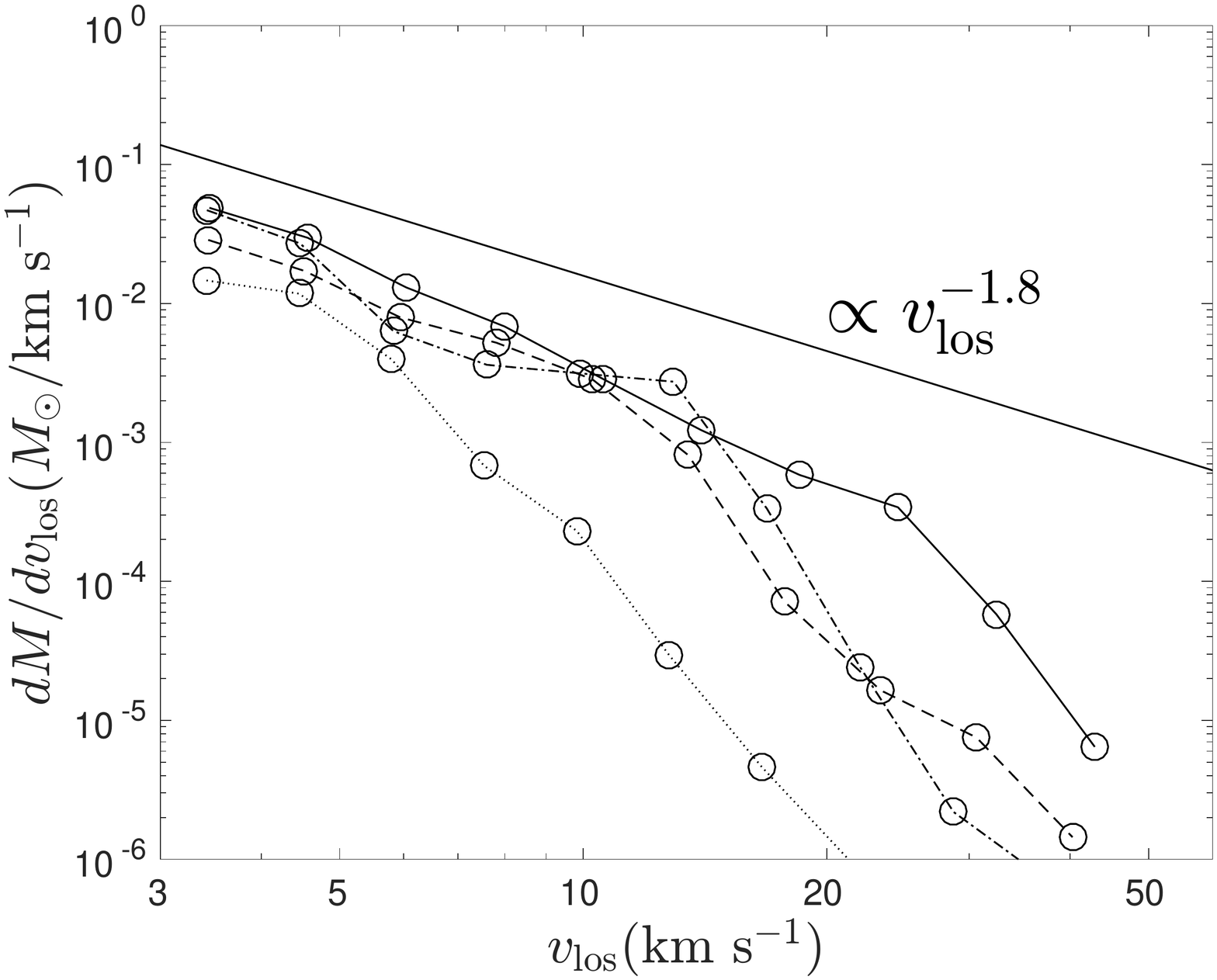}
\caption{Four examples of a broken power-law mass-velocity relation for simulated outflows. The line-of-sight velocity plotted is in the rest frame of each protostar. The low-velocity gas shows a velocity spectrum similar to $dM/dv_{\rm los} \propto v_{\rm los}^{-1.8}$ predicted in \citet{mat99} and seen in observations (e.g., \citealp{arc07}). 
The breaks in the power laws appear at 4 to $> 20$ km~s\e\ in the examples. The high-velocity gas has a much steeper slope, as seen in observations (e.g., \citealp{arc07}). 
See Section \ref{sec:outflow} for discussion.
\label{fig16}}
\end{figure}

\subsubsection{Results: The $M-v$ relation for protostellar outflows}

As the outflow sweeps up ambient gas, it decelerates. A diagnostic of the interaction of the outflow with the ambient gas is provided by the relation between the mass of the swept up gas and its velocity.
Observations of molecular outflows show that the mass-velocity ($M-v$) relation,
\beq
\frac{dM}{dv} \propto v^{-n},
\eeq
of a protostellar outflow is a broken power law \citep[see review by][and references therein]{arc07}. Observationally, there is a power law at low velocities with an index, $n$, of about 1 to 3 (usually around 1.8), and a much steeper power law at higher velocities, as steep as $n \sim 10$. \citet{mat99} proposed a simple self-similar model for the $M-v$ relation with $n \sim 1.8$ at low velocities. In Fig. \ref{fig16}, we provide several examples of outflow $M-v$ relations at the end of our simulation. 
To obtain the $M-v$ relation for a protostellar particle, we bin the mass in intervals of the line-of-sight velocity in the particle rest frame, $v_{\rm los}$, from 3 km~s$^{-1}$ up to the maximum velocity, in 10 bins. This procedure largely removes the background turbulent gas in the local frame.
A line with $dM/dv \propto v^{-1.8}$ is plotted in the figure for visual comparison. Fig. \ref{fig16} shows four examples of protostellar outflows in the simulation. The $M-v$ relation shows the typical broken power law relation that the low velocity part is close to $dM/dv \sim v^{-1.8}$ as predicted by \citet{mat99} and observed. Following this shallow power-law region, there is a region with a much steeper slope. From observations, the break could be anywhere from 3 to 30 km~s$^{-1}$ (although usually in the range 6-12 km~s$^{-1}$) \citep[e.g.][]{ric00,arc07}. In Fig. \ref{fig16}, we see that the breaks in the four example occur between 4 and $\sim$ 20~km~s$^{-1}$. It should be noted that in the simulation, most of the material above the break is in the jet. Because the jet is not well resolved in our simulation, the slope above the break in Fig \ref{fig16} is very approximate. In the simulation, protostars with a low break velocity have either a weak outflow, generally due to their youth, or an outflow with a large inclination angle with respect to the line of sight. \citet{mat99} predicted a single power law that terminates at a maximum velocity; in practice, this could appear as a very sharp break in the spectrum. Their results apply only to the swept-up material, not the protostellar winds and jets.

We do not specifically have a refinement criterion for outflows in the AMR simulation. Therefore, long outflows can span more than 3 different levels of refinement. To make sure the properties of the outflows are converged, we perform a convergence study on a pair of long outflows using a rectangular box enclosing the entire outflow region and refined to the highest level everywhere inside this box. The $M-v$ relation we obtained by this test is within the uncertainty of $\pm0.2$ in index $n$ when fitting the $M-v$ relation.
This implies that the outflow is converged even if it crosses several refinement levels.
The examples in Fig. \ref{fig16} are consistent with the observations, even though we are using a relatively simple outflow sub-grid model. Much higher resolution simulations are required to understand the actual driving mechanism for protostellar outflows.

\section{Conclusions}
\label{sec:conclusion}

Filamentary Infrared Dark Clouds (IRDCs) are embedded in GMCs and have a wide range of length scales. They are believed to be the precursors of massive stars and star clusters since massive gravitationally bound dense clumps and cores are observed to be forming inside them. Some of the clumps and cores have higher temperatures than the overall cloud, indicating the early stages of protostar formation. 

In this paper, we perform a high resolution zoom-in AMR simulation of star formation inside a long dark filament that formed in our large-scale ideal MHD IRDC simulation \citep{li15}. We study the properties of the protostellar cluster that forms in the filamentary cloud and its surrounding environment. The initial conditions are inherited from a large-scale simulation of an isothermal IRDC with a moderately strong magnetic field (\alfven Mach number of 1) and continuously driven supersonic turbulence with a sonic Mach number of 10. The gravitational collapse of this filamentary cloud started 350,000 years before this simulation begins. At the beginning of this simulation, the Truelove criterion \citep{tru97} has not been violated, so no sink particles have been formed. We include radiative feedback using flux-limited diffusion and protostellar outflow feedback using a sub-grid outflow model \citep{cun11}. We allow up to 6 levels of refinement in the $512^3$ base grid but limit the refinement to a 4.2 pc$^3$ region around the main cloud filament in order to reduce the computational cost of the radiative transfer calculation. The highest resolution achieved in this simulation is $\sim 28$ au. This simulation lasts about 350,000 years, or $1.65 \tffo$, where $\tffo$  is the free-fall time in dense gas at the beginning of the simulation (see Equation \ref{eq:tffo} for a precise definition).

At the end of the simulation, we have a total of 82 protostars created inside the 4.2 pc$^3$ zoom-in region. Since protostellar particles are allowed to be created only at the highest refined level, no protostellar particles are created outside the zoom-in region. From the 2-level IRDC simulation that provided the initial conditions for the present simulation, we know that the majority of the dense clumps and cores are located along the main dark cloud filament in the zoom-in region. Thus the star formation inside the zoom-in region is representative of the entire region. Here we summarize our main findings.

\begin{itemize}

\item[1.] {\it Spatial distribution of protostars.} The emerging protostars are aligned along the main filamentary cloud, as has been observed in dark cloud filaments \citep{zha15}.
Most of the protostars are aligned in a filamentary structure of width $\sim 0.1$ pc.
We find that there is no preferential star formation sequence along the filament, such as from one end to the other. The protostars form at any location in the filament where the density is high enough for gravitational collapse to occur. However, because of a collision with another filamentary cloud, the star formation rate increases significantly in the collision region, which is $\sim1$ pc long. At the end of the simulation, there are 51 (out of a total of 82) protostars in the collision region, about 6 times the number of protostars per unit length in the rest of the main filamentary cloud. The ratio of protostellar mass to total gas mass in the collision region is about 2.4 times that in the rest of this cloud.  This shows that
cloud-cloud collisions can have an important role in forming star clusters.

\item[2.] {\it Star formation efficiency of the cluster.} 
The SFE in the zoom-in region at the end of the simulation is 4.3 percent.
We find that the star formation efficiency inside the zoom-in region increases in time as SFE $\propto t^2$, similar to other numerical simulations \citep[e.g.][]{pad14,lee15}. For individual massive stars, a stellar mass $m_*\propto t^2$ is predicted by the Turbulent Core model \citep{mck03} for the observed typical density gradient in high-mass star forming regions, $\rho\propto r^{-1.5}$, and by the Murray-Chang model \citep{mur15} without the observational input.
Neither model includes the effects of protostellar feedback, so it is interesting that we find this result in a simulation that includes feedback, although it must be noted that no massive stars have yet formed in our simulation.
We find that the number of protostars in the simulation increased as $\caln_* \propto t^{1.47}$, indicating accelerated star formation, as seen in observations \citep{pal00}.

\item[3.] {\it Star formation rate per free-fall time.}
The dimensionless star formation rate per free-fall time, $\eff$, in the zoom-in region at the end of the simulation is 0.036 and 0.046 for the $A_K = 0.8$ and 0.1 mag contours, respectively, in the extinction map method. The theoretical value for $\eff$ (Equation \ref {eq:epsilon}), is $\efft=0.022$. These values $\eff$ are within the range of the measured rates of the galactic data samples in \citet{kru12a}.

\item[4.] {\it Protostellar accretion rates.}
The mean and median protostellar mass accretion rates in the simulation reach asymptotic values of $2.5 \times 10^{-6}$ and $1 \times 10^{-6}\, M_\odot$ yr$^{-1}$. If we exclude the protostars with masses $< 0.04 M_{\odot}$ because of inaccurate treatment of the formation of the first core, the mean and median mass accretion rates are 3.9 and $1.4 \times 10^{-6}\,M_\odot$  yr$^{-1}$, respectively. 
We confirm the results of \citet{pad14} that there is a significant dispersion in the accretion rates of individual protostars and among different protostars: 
We find that
the dispersion of the mean accretion rates of the simulated protostars is 0.54 dex, and the average dispersion of the accretion rates of individual stars is 0.51 dex. 
There are four protostars that have accretion rates exceeding $10^{-5}\, M_\odot$ yr$^{-1}$ for some period of time, and one of these protostars remains at or above this rate for its whole lifetime and becomes the most massive protostar ($\sim 3.1\, M_{\odot}$) at the end of the simulation. 

\item[5.] {\it Density profile of cloud cores around protostars.} \citet{mur15} study the density profile of protostellar cores using a spherically symmetric model. They find that the density profile is $\rho \propto r^{-3/2}$ inside the ``radius of influence," which is defined as the radius at which the enclosed gas mass equals the enclosed stellar mass. Numerical simulations by \citet{mur17} verify that the $\rho \propto r^{-3/2}$ profile is roughly time independent. 
We examine the density profiles of the cores around all our protostars and find that 85 per cent of the protostars have gas density profiles close to $\rho \propto r^{-3/2}$,
and for 70 per cent of these profile extends out to at least twice
the radius of influence. 
Thus, the outflows and radiative feedback included in our simulation do not lead to significant deviations from the $r^{-3/2}$ profile.

\item[6.] {\it Temperature structure around protostars.} As a result of the large accretion luminosity, the gas surrounding the protostars are heated up to higher temperature.
Most of the gas around the protostars has a temperature of $\ga 50$ K within 250 au and $\ga 20$ K within 1000 au.
The maximum is close to 100 K at the end of the simulation. There are small amounts of gas in the simulation at even higher temperature due to the shocked outflow from protostars.

\item[7.] {\it The protostellar mass function (PMF).} 
The PMF evolves in time and is in good agreement with the \citet{chab05} IMF at the end of the simulation; in particular, the
median mass is close to $0.20\,M_\odot$. The median accretion rate at late times in the simulation is about $1.0\times 10^{-6}\,M_\odot$~yr\e. The implied median star formation time is then
$t_{\rm SF}=m_{*,\rm med}/\dot m _{*,\rm med}\simeq 2\times 10^5$~yr, 
close to the mass-weighted median free-fall time.
Comparing with theoretical models of the PMF, we find that at the end of the simulation the 
PMF is close to the two-component turbulent core and two-component competitive accretion models \citep{mckoff10}, in which the protostar is assumed to accrete as in the \citet{shu77} solution at early times and as in the turbulent core or competitive accretion models, respectively, at late times. All three PMFs are also close to the Chabrier IMF. The theoretical PMF models have yet to include the large dispersion in accretion rates of individual protostars and of the mean accretion rates found by \citet{pad14} and in this work.

\item[8.] {Protostellar luminosities.}
We have calculated the protostellar luminosities, including the internal luminosity due to nuclear reactions and to Kelvin-Helmholz contraction and allowing for the energy required to dissociate and ionize the accreted gas.
In comparing with the observations of \citet{dun13}, we considered only sources with $L>0.04\,L_\odot$, 
the effective completeness limit of the data. 
The observed, simulated, and theoretical luminosities summarized in Table \ref{tab:luminosities} are in remarkably good agreement.
A significant amount of mass can be added to a protostar in FU Ori events \citep{ken90,har96}; \citet{off11} estimated that 25 per cent of the mass could be added this way. These events are so rare that none are apparent in the \citet{dun13} sample. Even
after subtracting the estimated luminosity associated with these unobserved events from the simulated values, we find that the mean and median luminosities remain quite close to the observed values.
As shown in Fig. \ref{fig11}, the shapes of the observed and simulated protostellar luminosity functions (PLFs) are similar. The dispersion of the simulated PLF is 0.76 dex, close to that of the observed PLF, 0.65 dex. 
It is also instructive to compare with the theoretically predicted luminosity given by Equation (\ref{eq:lacc}). For fiducial values of the parameters ($r_*=2.5\,R_\odot$, $m_*=0.25\,M_\odot$, $\dot m_*=2.5\times 10^{-6}\,M_\odot$~yr\e, and $f_{\rm acc,\, eff}=0.56$ including the effect of FU Ori outbursts), the theoretically predicted average accretion luminosity is $3.7\,L_\odot$ (Eq. \ref{eq:lacc}). This is about 2/3 of the observed and the simulated values, but the difference is well within the uncertainties in the values of the parameters. 

\item[9.] {\it The protostellar luminosity problem.} The classical protostellar luminosity problem is that the observed protostellar luminosities are significantly smaller than expected from simple theoretical estimates \citep{ken90}. The problem has been alleviated to some extent since modern observations show that the mean protostellar luminosity is about twice the value adopted by \citet{ken90}. Furthermore, \citet{ken90} compared the mode of the observed luminosity distribution with the mean of the theoretical one, which increased the discrepancy. If 25 per cent of the accretion energy is used to drive a protostellar outflow and 25 per cent is emitted in FU Ori outbursts that are too rare to be included in the observed sample, then theoretically predicted accretion luminosity (Eq. \ref{eq:lacc}) is 0.6-1.3 times the observed mean, so there is no protostellar luminosity problem. This conclusion agrees with that of \citet{off11} and \citet{pad14}. Our results extend those of \citet{off11} and {pad14} by showing that the inclusion of radiation and outflows does not alter the conclusion that the luminosity problem is resolved.

\item[10.] {\it Stellar multiplicity.} 
The companion fraction of the protostars at the end of the simulation matches reasonably well with observation over different ranges of separation to within a factor of 2. Because of the geometry of filamentary clouds, the multiplicity and companion fractions can change by up to a factor of 2 when they are observed along different directions.
From the results of the simulation, one can also determine the multiplicity based on whether the protostars 
in the system are gravitationally bound. We find that most of the multiple systems identified from 2D projections in the range 100-5000 au are not gravitationally bound, either because the kinetic energy is larger than the potential energy or because of the tidal force from nearby protostars. On the other hand,
80 per cent of the systems with projected separations in the range 100-400 au are bound; all of these are binaries.

\item[11.] {\it Outflows.} Using a sub-grid model for protostellar outflows, we simulate the properties of the observed outflows from protostars. Relatively collimated outflows, both straight and curved, are observed in the simulation. The largest extent of the outflows observed in the simulation is about 1 pc from the protostars.  Curved and unequal outflows are the results of interaction with nearby dense ambient gas. The mass-velocity relation of the outflows from the simulation is consistent with observations that show a broken power law relation, with the shallower power index around -1.8, as predicted in \citet{mat99} and observed \citep{arc07}. In our model, in which an outflow is launched parallel to the angular momentum of the protostar.
the orientations of the outflows are strongly correlated with the direction of the mean magnetic field.

\end{itemize}

\noindent
\section*{Acknowledgments}

We would like to thank Gaspard Duch\^{e}ne, Lee Hartmann, Norm Murray, and Stella Offner for helpful discussions on different topics in this paper. We also thank the referee for a number of helpful comments and suggestions on the paper. We particularly appreciate Michael Dunham's input on his published results.
Support for this research was provided by NASA through NASA ATP grant NNX13AB84G (RIK, CFM, and PSL), the US Department of Energy at the Lawrence Livermore National Laboratory under contract DE-AC52-07NA 27344 (RIK), and the NSF through grant AST-1211729 (CFM and RIK).  This work used the Extreme Science and Engineering Discovery Environment (XSEDE), which is supported by National Science Foundation grant number ACI-1053575, through the grant TG-MCA00N020, the computing resources provided by the NASA High-End Computing (HEC) Program through the NASA Advanced Supercomputing (NAS) Division at Ames Research Center, and the computing resources of the National Energy Research Scientific Computing Center, a DOE Office of Science User Facility supported by the Office of Science of the U.S. Department of Energy under Contract No. DE-AC02-05CH11231.

\label{lastpage}

\end{document}